\documentclass[12pt]{article}

\usepackage[dvips]{graphicx}
\usepackage{epsfig}
\usepackage{amsmath,amsfonts,amssymb,amsthm}
\usepackage{verbatim}
\usepackage{psfrag}
\usepackage{bm}
\usepackage{bbm}
\usepackage[square,comma,sort&compress,numbers]{natbib}
\usepackage{color}
\usepackage{slashed}

\usepackage{epsf,epsfig}
\usepackage{graphics}

\def\be{\begin{equation}}
\def\ee{\end{equation}}
\def\bea{\begin{eqnarray}}
\def\eea{\end{eqnarray}}

\begin{document}
\thispagestyle{empty}

\begin{flushright}
{
\small
TTK-10-50\\
}
\end{flushright}

\vspace{0.4cm}
\begin{center}
\Large\bf\boldmath
Leptogenesis: The Other Cuts
\unboldmath
\end{center}

\vspace{0.4cm}

\begin{center}
{Bj\"orn~Garbrecht}\\
\vskip0.3cm
{\it Institut f\"ur Theoretische Teilchenphysik und Kosmologie,\\ 
RWTH Aachen University,\\
D--52056 Aachen, Germany}\\
\end{center}

\vspace{1.3cm}
\begin{abstract}
\vspace{0.2cm}\noindent
For standard leptogenesis from the decay of singlet right-handed neutrinos,
we derive source terms for the lepton asymmetry
that are present in a finite density background
but absent in the vacuum. These arise from cuts through the vertex
correction to the decay asymmetry, where in the loop either
the Higgs boson and the right-handed
neutrino or the left-handed lepton and the right-handed neutrino are
simultaneously
on shell. We evaluate the source terms numerically and use them to calculate
the lepton asymmetry for illustrative points in parameter space, where
we consider only two right-handed neutrinos for simplicity.
Compared to calculations where only the standard cut through the
propagators of left-handed
lepton and Higgs boson is included, sizable corrections arise when
the masses of the right-handed neutrinos are of the same order,
but the new sources are found to be most relevant when the decaying
right-handed neutrino is heavier than the one in the loop. In that situation,
they can yield the dominant contribution to the lepton
asymmetry.

\end{abstract}

\newpage
\setcounter{page}{1}

%\newpage
\allowdisplaybreaks[2]

\section{Introduction}

Leptogenesis~\cite{Fukugita:1986hr} is often studied in parametric regimes
where the
masses of the right-handed neutrinos are either hierarchical or degenerate.
The hierarchical limit is particularly useful for gaining valuable
analytical insights
into the connections between leptogenesis and the observed
neutrino oscillations~\cite{Davidson:2002qv,Buchmuller:2002rq,Buchmuller:2003gz}.
To be specific, we discuss here the simple case of two right-handed neutrinos
$N_{1,2}$ with masses $M_{1,2}$ and Yukawa couplings $Y_{1,2}$ to
the Higgs and lepton doublets, {\it i.e.} a model as specified
in Ref.~\cite{Beneke:2010wd}.
The key simplification in the hierarchical limit, $M_1\ll M_2$,
is that the evolution of the lepton asymmetry during leptogenesis as a function of $z=M_1/T$ depends up to a proportionality factor only
on the ratio $M_1/(Y_1^2 m_{\rm Pl})$, which characterises the washout strength.
Here, $T$ is the temperature and $m_{\rm Pl}$ is the Planck mass.
The remaining proportionality factor characterising the amount of
$CP$ violation is
${\rm Im}[Y_1^2 {Y_2^*}^2]M_1/M_2$, up to corrections of order
$M_1^3/M_2^3$, which one neglects in the hierarchical limit.
On the other hand, mass-degenerate right-handed neutrinos lead to
resonant leptogenesis~\cite{Flanz:1994yx,Flanz:1996fb,Pilaftsis:1997jf,Covi:1996wh,Pilaftsis:2003gt}.
This corresponds to
a phenomenologically attractive scenario, since the decay asymmetry
is enhanced, which allows for lower temperatures at which leptogenesis
takes place. Thus, the production of unwanted relics, most notoriously of
gravitinos within supersymmetric models~\cite{Khlopov:1984pf,Ellis:1984eq,Ellis:1984er}, along with the lepton asymmetry
asymmetry can be avoided. A lower energy scale might also give rise to new experimentally accessible signals
connected with leptogenesis, see {\it e.g.}~\cite{Pilaftsis:2005rv,Blanchet:2009bu,Blanchet:2009kk}.

\begin{figure}[t!]
\begin{center}
%\begin{tabular}{c}
\epsfig{file=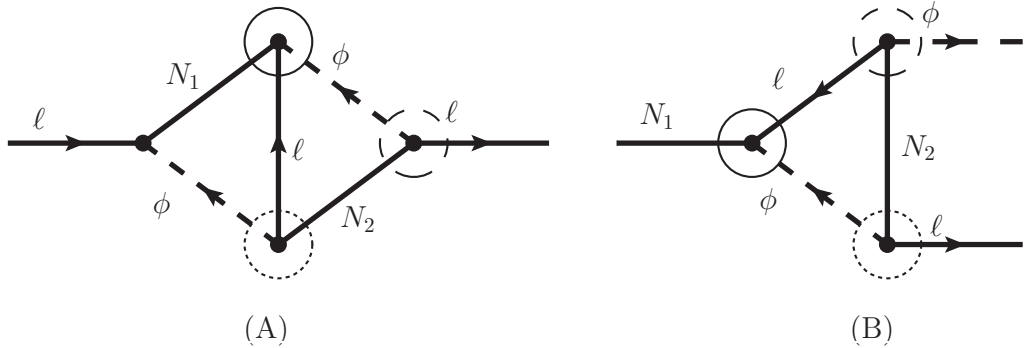,width=14.cm}
%\hskip.5cm
\end{center}
\caption{
\label{fig:doublediamond}
Diagram~(A) represents the
vertex correction ${\Sigma\!\!\!/}^{{\rm v}>}_{\ell}$ to the lepton self-energy in the CTP formalism.
Diagram~(B) is the subdiagram of (A) that accounts for the decays
and inverse decays of the out-of-equilibrium particle $N_1$.
We indicate the various cuts that arise from demanding that the cut particles
in the loop are on shell. The solid circle represents the standard cut
through $\{\ell,\phi\}$ that
is the only contribution in the vacuum or when finite density effects are neglected. The dashed cut
is the contribution for off-shell $\ell$ through $\{\phi,N_2\}$, the dotted cut the the contribution for
off-shell $\phi$ through $\{\ell,N_2\}$.
}
\end{figure}

However, the origin of the masses of the right-handed neutrinos is yet 
unknown and their masses may well be neither hierarchical nor degenerate.
When the hierarchical limit $M_1\ll M_2$ no longer applies,
it is well known that the decay asymmetry of $N_1$ is not simply proportional
to $M_1/M_2$ with negligible corrections, as can be verified by inspection
of the vertex and wave-function contributions to the decay asymmetry of
$N_1$~\cite{Fukugita:1986hr,Covi:1996wh}.
In the finite-temperature background, there are additional corrections due
to new cuts. While in the vacuum background, the $CP$-violating 
contribution from the vertex function arises exclusively from the cut through
$\{\ell,\phi\}$, where the
internal lepton and Higgs boson are on-shell,
at finite 
temperature also the two other possible cuts
through $\{\ell,N_2\}$ or $\{\phi,N_2\}$
in the vertex diagram contribute,
{\it cf.} Figure~\ref{fig:doublediamond}~(B). This is because in
the finite-temperature background, the cut-particles do not need to correspond
to stimulated (suppressed) emission processes for bosons (fermions), but
they can also correspond to absorption processes of particles from the plasma.
The presence of these cut contributions has been mentioned in
Ref.~\cite{Giudice:2003jh}. However, by now, neither
analytical expressions for these
terms have been derived nor have these been evaluated numerically in order to
compute effective decay asymmetries or the resulting lepton asymmetry.
These  are the main goals of the present work.

The new cut contributions are a finite density effect, and a powerful
method of describing out-of-equilibrium field theory is given by the
Schwinger-Keldysh Closed-Time-Path (CTP) formalism~\cite{Schwinger:1960qe,Keldysh:1964ud}. This approach has been
applied to leptogenesis and has resulted in some recent activity
which we build upon within the present work~\cite{Buchmuller:2000nd,De Simone:2007rw,Garny:2009rv,Garny:2009qn,Garny:2010nj,Anisimov:2010aq,Beneke:2010dz,Garny:2010nz}. Main advantages of the CTP 
approach to leptogenesis over the conventional description by semi-classical Boltzmann equations may be seen in the absence of the need of an explicit
subtraction procedure for real intermediate states (RIS)~\cite{Kolb:1979qa}
and in the systematic
treatment of finite-density corrections.

Within the CTP approach, the vertex diagram in Figure~\ref{fig:doublediamond}~(B)
appears as a subdiagram in the self-energy Figure~\ref{fig:doublediamond}~(A),
which is a contribution
to the lepton self-energy. This self-energy in turn enters the collision term of
the Kadanoff-Baym equation for the lepton, that can be reduced to a kinetic equation
which describes the gain and the loss and therefore the time evolution of the
lepton number density. In order to simplify the collision term to a
manageable form, it is useful to substitute equilibrium propagators for
$\ell$ and $\phi$ and to employ Kubo-Martin-Schwinger (KMS) relations.
The present work relies strongly on Ref.~\cite{Beneke:2010wd},
where these simplification 
strategies are explained and justified in detail and where also
many definitions and quantities that we use here are introduced.
An additional simplifying assumption that we introduce here, but that will
not hold true in general, is that also $N_2$ is maintained in equilibrium
({\it e.g.} through interactions with an additional lepton flavour), such that
$N_1$ is the only out-of-equilibrium particle. We leave a study of the situation
where more than one of the right-handed neutrinos deviates from equilibrium
to future work.

Provided the initial distribution function for $N_1$ is thermal, as we
assume in the present work, the main contributions to the lepton asymmetry
occur at times when the value of $z=M_1/T$ is in the range of around one up
to a few, for an in-detail discussion, see Ref.~\cite{Buchmuller:2004nz}. This is because at these temperatures, $N_1$ becomes non-relativistic
and therefore deviate from equilibrium. As a consequence,
we anticipate the new cuts to be important under the following conditions:
\begin{itemize}
\item
Since the effect is due to the finite densities, $M_2$ should not be
much larger than $M_1$, because otherwise the new contributions are
Maxwell-suppressed, just as the equilibrium distribution functions for energies
of order $M_2$
much larger than $T$.
\item
The new effects will be most pronounced in case leptogenesis occurs
at comparably low values for $z$, when it is around one.
This is the case in the transitional regime between
strong and weak washout. In such a situation, the finite density effects that include
the contributions from the new cuts, take their largest relevance.
\item
A loophole to
these arguments are situations where $M_1>M_2$. (We denote within this work
by $N_1$ the neutrino that deviates from equilibrium, whereas $N_2$ is assumed
to be very close to equilibrium at the time relevant for leptogenesis. This
definition differs from what is commonly used in the literature, where
$N_1$ corresponds to the lightest right-handed neutrino,
{\it cf.} Refs~\cite{DiBari:2005st,Engelhard:2006yg,Bertuzzo:2010et}.
In this case, the finite density
effects from the $N_2$ are also important when leptogenesis occurs at larger
values of $z$, since the distribution of $N_2$ can still be unsuppressed.
\end{itemize}

The plan of this paper is as follows:
In Section~\ref{sec:vertex}, we give the expressions for the $CP$-violating
source terms that bias the lepton number and that are valid for finite
ratios of $M_1/M_2$. For the standard cut through
$\{\ell, \phi\}$, we quote the result from Ref.~\cite{Beneke:2010wd}, while for 
the new cuts through $\{\ell, N_2\}$ and $\{\phi, N_2\}$, respectively,
we derive new expressions. In Section~\ref{sec:examples}, we present the first 
numerical results for the cut through
$\{\ell, \phi\}$ at finite $M_1/M_2$ and finite density, as well as for
the new cuts through $\{\ell, N_2\}$ and $\{\phi, N_2\}$. We first define
and evaluate expressions for the effective $CP$ violation for several ratios
of $M_1/M_2$. Then, we solve the kinetic equations for the lepton asymmetry for
the same values of $M_1/M_2$ and specific illustrative choices of $Y_1$,
$Y_2$. The results for the lepton asymmetry are compared with the
effective $CP$ violation as well as with what is stated above on the parametric
regions where we anticipate the new cut contributions to be most relevant. Within flavoured leptogenesis, a new cut
in the wave-function of the lepton $\ell$ contributes to the asymmetry. A 
rough estimate of this effect is presented in Section~\ref{sec:flavour},
and it is found to be negligibly small. We summarise and conclude in
Section~\ref{sec:conclusions}.

\section{Thermal Vertex Function}
\label{sec:vertex}

\subsection{Vertex Self Energy and Collision Term}

Within the present work, we extend the results of Ref.~\cite{Beneke:2010wd}.
There, the kinetic
evolution equation for the lepton asymmetry is expressed as
\begin{align}
\label{kin:eq}
\frac{d}{d\eta}\left(
n_\ell-\bar n_\ell
\right)=W+S
\,,
\end{align}
where $\eta$ is the conformal time in the Friedmann background,
$W$ the washout term, $S$ the source 
term and $n_\ell$ ($\bar n_\ell$) the comoving (anti-)lepton number density.
In the radiation-dominated Universe, the scale factor is given by
$a=a_{\rm R}\eta$, where $a_{\rm R}$ is an arbitrary constant. The washout
term $W$ is discussed in detail in Ref.~\cite{Beneke:2010wd}. It encompasses
the tree-level decay and inverse decay processes of the right-handed neutrino.
The source term accounts for the $CP$-violating loop effects. It decomposes as
\begin{align}
S=\int\frac{d^3 k}{(2\pi)^3}\left[
{\cal C}_\ell^{\rm wf}(\mathbf{k})
+{\cal C}_\ell^{\rm v}(\mathbf{k})
\right]\,,
\end{align}
where ${\cal C}_\ell^{\rm wf}$ is the wave-function contribution to the
collision term, as it is given in Ref.~\cite{Beneke:2010wd}. In this work,
we are primarily concerned with the vertex
contribution ${\cal C}_\ell^{\rm v}$ and the new 
corrections that it acquires when compared with Ref.~\cite{Beneke:2010wd}.
This term can be expressed in the usual Kadanoff-Baym form
\begin{align}
\label{collision:KB}
{\cal C}^{X}_\ell(\mathbf k)=
\int\frac{dk^0}{2\pi}
{\rm tr}\left[
{\rm i}{\Sigma\!\!\!/}^{X>}_\ell(k)P_{\rm L}{\rm i}S_\ell^<(k)
-{\rm i}{\Sigma\!\!\!/}^{X<}_\ell(k)P_{\rm L}{\rm i}S_\ell^>(k)
\right]\,,
\end{align}
for $X\equiv{\rm v}$.
The Wightman self-energy ${\Sigma\!\!\!/}^{{\rm v}>}$ has the diagrammatic
representation of Figure~\ref{fig:doublediamond}~(A). It is given
by~\cite{Beneke:2010wd}
\begin{align}
\label{Sigma:v}
{\rm i}{\Sigma\!\!\!/}^{{\rm v}>}_{\ell}(k)
=&-{Y_i^*}^2 Y_j^2
\int\frac{d^4p}{(2\pi)^4}
\frac{d^4q}{(2\pi)^4}
\\\notag
\Big\{
&{\rm i}S_{Ni}^>(-p)
C\left[{\rm i}S_\ell^T(p+k+q)\right]^tC^\dagger
{\rm i}S^T_{Nj}(-q)
{\rm i}\Delta_\phi^<(-p-k)
{\rm i}\Delta_\phi^T(-q-k)
\\\notag
-&{\rm i}S_{Ni}^{\bar T}(-p)
C\left[{\rm i}S_\ell^<(p+k+q)\right]^tC^\dagger
{\rm i}S^T_{Nj}(-q)
{\rm i}\Delta_\phi^<(-p-k)
{\rm i}\Delta_\phi^<(-q-k)
\\\notag
-&{\rm i}S_{Ni}^>(-p)
C\left[{\rm i}S_\ell^>(p+k+q)\right]^tC^\dagger
{\rm i}S^>_{Nj}(-q)
{\rm i}\Delta_\phi^{\bar T}(-p-k)
{\rm i}\Delta_\phi^T(-q-k)
\\\notag
+&{\rm i}S_{Ni}^{\bar T}(-p)
C\left[{\rm i}S_\ell^{\bar T}(p+k+q)\right]^tC^\dagger
{\rm i}S^>_{Nj}(-q)
{\rm i}\Delta_\phi^{\bar T}(-p-k)
{\rm i}\Delta_\phi^<(-q-k)
\Big\}\,,
\end{align}
where we sum over the indices $i,j$.
The self-energy ${\Sigma\!\!\!/}^{{\rm v}<}$ follows when applying the
replacements $<\leftrightarrow >$ and $T\leftrightarrow \bar T$. For the
propagators $S_\ell$, $\Delta_\phi$ and $S_{Ni}$,
we use the zero-width approximations as written down
in Ref.~\cite{Beneke:2010wd}.

To be specific, let us now consider the term with $i=1$ and $j=2$.
The $CP$-violating contributions from the decays and inverse decays of $N_1$
arise when two out of the three propagators
${\rm i}S_\ell(p+k+q)$, ${\rm i}\Delta_\phi(-p-k)$ and
${\rm i}S_{N2}(-q)$ are on shell. These are the cut propagators.
As a consequence of this, the third 
propagator is off shell. Since the $>$ and $<$ propagators
are purely on shell, it follows that only terms where the off-shell
propagator is of the time-ordered $T$ or anti-time-ordered $\bar T$ type
contribute.
The collision term~(\ref{collision:KB}) can therefore be split
into the portions
\begin{align}
{\cal C}^{{\rm v}}_\ell
={\cal C}^{{\rm v}\ell\phi}_\ell
+{\cal C}^{{\rm v}\phi N2}_\ell
+{\cal C}^{{\rm v}\ell N2}_\ell\,,
\end{align}
where the superscripts indicate through which two of the loop propagators
in the vertex correction the cut goes.
Likewise, we decompose the vertex contribution to the source term as
\begin{align}
\label{S:v}
S^{\rm v}=S^{{\rm v}\ell\phi}+S^{{\rm v}\phi N2}+S^{{\rm v}\ell N2}\,.
\end{align}

Within the present work, we make the simplifying assumption that $N_2$ is in 
thermal equilibrium at all times,
whereas $N_1$ is in equilibrium initially but then deviates from equilibrium
when it becomes non-relativistic. We note that in general, both $N_1$ and $N_2$
will deviate from equilibrium at the same time. In fact, situations are conceivable
where the initial abundance of $N_2$ is zero and $Y_2$ is so small that $N_2$ does
not equilibrate before becoming non-relativistic. Therefore, the equilibrium deviation
of $N_2$ can be large when compared to $N_1$, and inverse decays of $N_2$ may
largely enhance the lepton asymmetry. For now, leave this interesting possibility for
future work and note that $N_2$ may be maintained in equilibrium by a stronger coupling
to a different lepton flavour within which no asymmetry is produced.

\subsection{Cut through $\{\ell,\phi\}$}

The term ${\cal C}^{{\rm v}\ell\phi}_\ell$ arises from those terms within
${\rm i}{\Sigma\!\!\!/}^{{\rm v}<,>}_{\ell}$, Eq.~(\ref{Sigma:v}), where
the propagators
${\rm i}S_{N2}^{T,\bar T}$ occur. It can be further simplified when
approximating the distribution functions of $\ell$ and $\phi$ by Fermi-Dirac
and, respectively, Bose-Einstein equilibrium distributions
$f^{\rm eq}_\ell(\mathbf k)$ and $f^{\rm eq}_\phi(\mathbf k)$. Accounting for
the fact that $\ell$ deviates from equilibrium only leads to higher order corrections
in the gradient expansion, as it is explained in Ref.~\cite{Beneke:2010wd}.
The gradient expansion
corresponds to an expansion in powers of $Y_1^2$ or equivalently
$H/T$, where $H$ denotes the Hubble rate.

The contribution of the cut through $\{\ell,\phi\}$
to the source term can be factorised into~\cite{Beneke:2010wd}
\begin{align}
\label{GammaV}
\Gamma^{\ell \phi}_\mu(k,p^{\prime\prime};M_1,M_2)
=&\int
\frac{d^3 k^\prime}{(2\pi)^3 2|\mathbf k^\prime|}
\frac{d^3 k^{\prime\prime}}{(2\pi)^3 2|\mathbf k^{\prime\prime}|}
(2\pi)^4\delta^4(k-k^\prime-k^{\prime\prime})\,
k_\mu^\prime\,\frac{M_1 M_2}{(k^\prime-p^{\prime\prime})^2-M_2^2}
\\\notag
&\times
\left[
1-f^{\rm eq}_\ell(\mathbf k^\prime)+f^{\rm eq}_\phi(\mathbf k^{\prime\prime})
\right]\,
\end{align}
and
\begin{align}
V^{\ell \phi}(k,M_1,M_2)
=&\int\frac{d^3 p^\prime}{(2\pi)^3 2|\mathbf p^\prime|}
\frac{d^3 p^{\prime\prime}}{(2\pi)^3 2|\mathbf p^{\prime\prime}|}
(2\pi)^4\delta^4(k-p^\prime-p^{\prime\prime})\,
{p^\prime}^\mu\,\Gamma^{\ell \phi}_\mu(k,p^{\prime\prime};M_1,M_2)
\\\notag
&\times
\left[
1-f^{\rm eq}_\ell(\mathbf p^\prime)+f^{\rm eq}_\phi(\mathbf p^{\prime\prime})
\right]
\end{align}
with the result
\begin{align}
\label{coll:vertex:final}
S^{{\rm v}\ell \phi}=\int\frac{d^3 p^\prime}{(2\pi)^3}\,
{\cal C}_\ell^{{\rm v}\ell \phi}(\mathbf p^\prime)
=4\,{\rm Im}[Y_1^2 {Y_2^*}^2]\int \frac{d^3 k}{(2\pi)^3 2 \sqrt{{\mathbf k}^2+M_1^2}}\,
\delta f_{N1}(\mathbf k) \,V^{\ell \phi}(k;M_1,M_2)
\,.
\end{align}
The term $\delta f_{N1}(\mathbf k)$ denotes the deviation of $f_{N1}(\mathbf k)$ from
the equilibrium Fermi-Dirac distribution,
$\delta f_{N1}(\mathbf k)=f_{N1}(\mathbf k)-f^{\rm eq}_{N1}(\mathbf k)$.

In Ref.~\cite{Beneke:2010wd}, this result has been evaluated in the hierarchical
limit $M_1/M_2\ll 1$,
which has the virtue that, as it is explained in the Introduction, the evolution of the lepton
number density is independent of $M_2$ up to
an overall proportionality. Furthermore, when $M_1/M_2\ll 1$, the collision 
term~(\ref{coll:vertex:final}) can be reduced analytically to a one-dimensional integral,
which can easily be evaluated numerically. For the present work, we numerically
evaluate a multi-dimensional integral for $S^{{\rm v}\ell \phi}$ that remains when exploiting
the $\delta$-functions. We note that also within the related 
articles~\cite{Garny:2009rv,Garny:2010nj}, all
source terms have been evaluated in the hierarchical limit, such that here, we
present the first quantitative results for effects of finite $M_1/M_2$ in a finite-density
background.

\subsection{Cut through $\{\phi,N_2\}$}

Now the cut goes through the internal $N_2$ and $\phi$ lines
of Figure~\ref{fig:doublediamond}~(A), and the internal $\ell$
is off-shell. The according contributions to the collision term are
\begin{align}
{\cal C}_\ell^{{\rm v }\phi N2}(\mathbf k)
=&-{Y_i^*}^2 Y_j^2 \int \frac{dk^0}{2\pi}
\frac{d^4 p}{(2\pi)^4}\frac{d^4q}{(2\pi)^4}
\\\notag
{\rm tr}\Big\{&
{\rm i}S_\ell^<(k)
{\rm i}S_{Ni}^>(-p)
C\left[S_\ell^T(p+k+q)\right]^tC^\dagger
{\rm i}S_{Nj}^T(-q)
{\rm i}\Delta_\phi^<(-p-k)
{\rm i}\Delta_\phi^{T}(-q-k)
\\\notag
-&
{\rm i}S_\ell^<(k)
{\rm i}S_{Ni}^{\bar T}(-p)
C\left[S_\ell^T(p+k+q)\right]^tC^\dagger
{\rm i}S_{Nj}^>(-q)
{\rm i}\Delta_\phi^{\bar T}(-p-k)
{\rm i}\Delta_\phi^{<}(-q-k)
\\\notag
+&
{\rm i}S_\ell^>(k)
{\rm i}S_{Ni}^<(-p)
C\left[S_\ell^T(p+k+q)\right]^tC^\dagger
{\rm i}S_{Nj}^{\bar T}(-q)
{\rm i}\Delta_\phi^>(-p-k)
{\rm i}\Delta_\phi^{\bar T}(-q-k)
\\\notag
-&
{\rm i}S_\ell^>(k)
{\rm i}S_{Ni}^{T}(-p)
C\left[S_\ell^T(p+k+q)\right]^tC^\dagger
{\rm i}S_{Nj}^<(-q)
{\rm i}\Delta_\phi^{T}(-p-k)
{\rm i}\Delta_\phi^{>}(-q-k)
\Big\}\,.
\end{align}
% \begin{align}
% {\cal C}_\ell^{{\rm v }\ell}
% =&-{Y_i^*}^2 Y_j^2 \int \frac{dk^0}{2\pi}
% \frac{d^4 p}{(2\pi)^4}\frac{d^4q}{(2\pi)^4}
% \\\notag
% \frac 12{\rm tr}\Big\{&
% \left({\rm i}S_\ell^<(k)-{\rm i}S_\ell^>(k)\right)
% {\rm i}S_{Ni}^{>}(-p)
% {\rm i}\Delta_\phi^{<}(-p-k)
% C\left[S_\ell^T(p+k+q)\right]^tC^\dagger
% {\rm i}S_{Nj}^<(-q)
% {\rm i}\Delta_\phi^{>}(-q-k)
% \\\notag
% -&
% \left({\rm i}S_\ell^<(k)-{\rm i}S_\ell^>(k)\right)
% {\rm i}S_{Ni}^{<}(-p)
% {\rm i}\Delta_\phi^{>}(-p-k)
% C\left[S_\ell^T(p+k+q)\right]^tC^\dagger
% {\rm i}S_{Nj}^>(-q)
% {\rm i}\Delta_\phi^{<}(-q-k)
% \Big\}
% \end{align}

As explained above, we assume here that
besides $\ell$ and $\phi$,
$N_2$ is in equilibrium, such that its number density is given by
the Fermi-Dirac distribution function
$f^{\rm eq}_{N2}(\mathbf k)$. This allows us to apply KMS relations
and two further replacements, which become identities under the integrals
({\it cf.} Ref.~\cite{Beneke:2010wd}, where this is explained in more detail):
\begin{align}
{\rm i}S_{Ni}^{T,\bar T}(-p)
{\rm i}\Delta_\phi^{T\bar T}(-p-k)
&\to
\frac 12
\left[
{\rm i}S_{Ni}^{<}(-p)
{\rm i}\Delta_\phi^{>}(-p-k)
+
{\rm i}S_{Ni}^{>}(-p)
{\rm i}\Delta_\phi^{<}(-p-k)
\right]\,,
\\\notag
{\rm i}S_{Nj}^{T,\bar T}(-q)
{\rm i}\Delta_\phi^{T,\bar T}(-q-k)
&\to
\frac 12
\left[
{\rm i}S_{Nj}^{<}(-q)
{\rm i}\Delta_\phi^{>}(-q-k)
+
{\rm i}S_{Nj}^{>}(-q)
{\rm i}\Delta_\phi^{<}(-q-k)
\right]
\,.
\end{align}
The result of these simplifications is
\begin{align}
{\cal C}_\ell^{{\rm v }\phi N2}(\mathbf k)
=&-[{Y_1^*}^2 Y_2^2 -Y_1^2{Y_2^*}^2]\int \frac{dk^0}{2\pi}
\frac{d^4 p}{(2\pi)^4}\frac{d^4q}{(2\pi)^4}
\\\notag
\frac 12{\rm tr}\Big\{&
\Big[
{\rm i}S_\ell^<(k)
{\rm i}S_{N1}^{>}(-p)
C\left[S_\ell^T(p+k+q)\right]^tC^\dagger
{\rm i}\Delta_\phi^{<}(-p-k)
\\\notag
-&
{\rm i}S_\ell^>(k)
{\rm i}S_{N1}^{<}(-p)
C\left[S_\ell^T(p+k+q)\right]^tC^\dagger
{\rm i}\Delta_\phi^{>}(-p-k)
\Big]
\\\notag
\times&
\Big[
{\rm i}S_{N2}^<(-q)
{\rm i}\Delta_\phi^{>}(-q-k)
-
{\rm i}S_{N2}^>(-q)
{\rm i}\Delta_\phi^{<}(-q-k)
\Big]
\Big\}\,.
\end{align}

% \begin{align}
% \Gamma_\mu^\ell(p,p^\prime;M_1,M_2)=&
% 2\int\frac{d^3 k^{\prime\prime}}{(2\pi)^3 2|\mathbf k^{\prime\prime}|}
% \frac{d^3 k^\prime}{(2\pi)^3 2 \sqrt{{\mathbf k^\prime}^2+M_2^2}}
% (2\pi)^4\delta^4(k^\prime-k^{\prime\prime}-p^\prime)
% \\\notag
% \times&\frac{(p+k^{\prime\prime})_\mu}{(p+k^{\prime\prime})^2}M_1M_2
% \left[
% f_\phi^{\rm eq}(\mathbf k^{\prime\prime})+f_{N2}(\mathbf k^\prime)
% \right]
% \end{align}
% 
% \begin{align}
% V^\ell(p;M_1,M_2)=&
% \int\frac{d^3 p^\prime}{(2\pi)^32|\mathbf p^\prime|}
% \frac{d^3 p^{\prime\prime}}{(2\pi)^3|\mathbf p^{\prime\prime}|}
% (2\pi)^4\delta^4(p-p^\prime-p^{\prime\prime})
% {p^\prime}^\mu\Gamma_\mu^\ell
% \left[
% 1-f_\ell(p^\prime)+f_\phi(p^{\prime\prime})
% \right]
% \end{align}
% 
% \begin{align}
% S^{{\rm v}\ell}=4{\rm Im}[Y_1^* Y_2^*]
% \int\frac{d^3 p}{(2\pi)^3 2 \sqrt{{\mathbf p}^2+M_1^2}}
% \delta f_{N1}({\mathbf p}) V^\ell(p;M_1,M_2)
% \end{align}
The final contribution to the source term,
\begin{align}
S^{{\rm v}\phi N2}
=\int\frac{d^3 p^\prime}{(2\pi)^3}
{\cal C}_\ell^{{\rm v }\phi N2}(\mathbf p^\prime)\,,
\end{align}
then follows when substituting the explicit forms of the propagators
(as they can be found in Ref.~\cite{Beneke:2010wd}) as
\begin{align}
\notag
S^{{\rm v}\phi N2}=&4{\rm Im}[{Y_1}^2 {Y_2^*}^2]\int\frac{d^3 p^\prime}{(2\pi)^32|\mathbf p^\prime|}
\frac{d^3 p}{(2\pi)^3 2\sqrt{\mathbf p^2+M_1^2}}
\frac{d^3 p^{\prime\prime}}{(2\pi)^3|\mathbf p^{\prime\prime}|}
\frac{d^3 k^{\prime\prime}}{(2\pi)^3 2|\mathbf k^{\prime\prime}|}
\frac{d^3 k^\prime}{(2\pi)^3 2 \sqrt{{\mathbf k^\prime}^2+M_2^2}}
\\
\label{S:ell}
\times&
(2\pi)^4\delta^4(p-p^\prime-p^{\prime\prime})
(2\pi)^4\delta^4(k^\prime-k^{\prime\prime}-p^\prime)
{p^\prime}^\mu\frac{(p+k^{\prime\prime})_\mu}{(p+k^{\prime\prime})^2}M_1M_2
\\\notag
\times&
\delta f_{N1}({\mathbf p})
\left[
1-f^{\rm eq}_\ell(\mathbf p^\prime)+f^{\rm eq}_\phi(\mathbf p^{\prime\prime})
\right]
\times
\left[
-f_\phi^{\rm eq}(\mathbf k^{\prime\prime})-f^{\rm eq}_{N2}(\mathbf k^\prime)
\right]\,.
\end{align}
To our knowledge,
this is the first report of a result for a source term for leptogenesis that is present at
finite density but absent in the limit of a vacuum background.
We remark that the last factor results from the expression
$-f_\phi^{\rm eq}-f^{\rm eq}_{N2}=-[1+f_\phi^{\rm eq}]f^{\rm eq}_{N2}-f_\phi^{\rm eq}[1-f^{\rm eq}_{N2}]$. 
This
could also be guessed starting from a hypothetical loop factor when the decay
$\ell \to \phi N_2$ was kinematically allowed,
$[1+f_\phi^{\rm eq}-f^{\rm eq}_{N2}]=[1+f_\phi^{\rm eq}][1-f^{\rm eq}_{N2}]+f_\phi^{\rm eq}f^{\rm eq}_{N2}$,
and applying the replacements $-f^{\rm eq}_{N2}\leftrightarrow[1-f^{\rm eq}_{N2}]$. This argument may be 
considered as a consistency check for our derived
result~(\ref{S:ell}).

It is important to
notice that for $M_2\gg T$, the source term~(\ref{S:ell}) is strongly Maxwell suppressed.
While in such a situation, $f^{\rm eq}_{N2}(\mathbf k^\prime)$ is always suppressed because of
the large mass of $N_2$, the energy-momentum conserving $\delta$-functions always imply
that then at least one of the momenta $\mathbf p$ or $\mathbf k^{\prime\prime}$ is
much larger than $T$, such that $f_\phi^{\rm eq}(\mathbf k^{\prime\prime})$
or $\delta f_{N1}({\mathbf p})$ are suppressed as well.

\subsection{Cut through $\{\ell,N_2\}$}

We finally consider the cuts through the propagators $N_2$ and $\ell$  within the 
loop and 
take $\phi$ to be off-shell. The contribution to the collision term is
\begin{align}
{\cal C}_\ell^{{\rm v }\ell N_2}(\mathbf k)
=&-{Y_i^*}^2 Y_j^2 \int \frac{dk^0}{2\pi}
\frac{d^4 p}{(2\pi)^4}\frac{d^4q}{(2\pi)^4}
\\\notag
{\rm tr}\Big\{&
{\rm i}S_\ell^<(k)
{\rm i}S_{Ni}^>(-p)
C\left[S_\ell^>(p+k+q)\right]^tC^\dagger
{\rm i}S_{Nj}^>(-q)
{\rm i}\Delta_\phi^T(-p-k)
{\rm i}\Delta_\phi^{T}(-q-k)
\\\notag
-&
{\rm i}S_\ell^<(k)
{\rm i}S_{Ni}^{\bar T}(-p)
C\left[S_\ell^{\bar T}(p+k+q)\right]^tC^\dagger
{\rm i}S_{Nj}^>(-q)
{\rm i}\Delta_\phi^{T}(-p-k)
{\rm i}\Delta_\phi^{<}(-q-k)
\\\notag
+&
{\rm i}S_\ell^>(k)
{\rm i}S_{Ni}^<(-p)
C\left[S_\ell^<(p+k+q)\right]^tC^\dagger
{\rm i}S_{Nj}^{<}(-q)
{\rm i}\Delta_\phi^T(-p-k)
{\rm i}\Delta_\phi^{\bar T}(-q-k)
\\\notag
-&
{\rm i}S_\ell^>(k)
{\rm i}S_{Ni}^{T}(-p)
C\left[S_\ell^T(p+k+q)\right]^tC^\dagger
{\rm i}S_{Nj}^<(-q)
{\rm i}\Delta_\phi^{T}(-p-k)
{\rm i}\Delta_\phi^{>}(-q-k)
\Big\}\,.
\end{align}
Again, we substitute for $S_{N2}$, $S_\ell$ and $S_\phi$ the equilibrium
propagators and employ KMS relations. Under the integrals,
we make the replacements
\begin{align}
&{\rm i}S_{Ni}^{T,\bar T}(-p)
{\rm i}C\left[S_\ell^{T,\bar T}(p+k+q)\right]^tC^\dagger
\\\notag
\to
&
\frac12
\left[
{\rm i}S_{Ni}^{>}(-p)
{\rm i}C\left[S_\ell^{>}(p+k+q)\right]^tC^\dagger
+
{\rm i}S_{Ni}^{<}(-p)
{\rm i}C\left[S_\ell^{<}(p+k+q)\right]^tC^\dagger
\right]\,.
\end{align}
Furthermore, it is useful to notice that when substituting the equilibrium
propagator for $\ell$,
\begin{align}
\nonumber
&
{\rm i}S_\ell^<(k)
{\rm i}S_{Ni}^>(-p)
C\left[S_\ell^>(p+k+q)\right]^tC^\dagger
{\rm i}S_{Nj}^>(-q)
\\\notag
-&
{\rm i}S_\ell^>(k)
{\rm i}S_{Ni}^<(-p)
C\left[S_\ell^<(p+k+q)\right]^tC^\dagger
{\rm i}S_{Nj}^<(-q)
\end{align}
is odd under the exchange $k^0,p^0,q^0\to -k^0,-p^0,-q^0$,
while ${\rm Im}[{\rm i}\Delta_\phi^{T,\bar T}(-q-k)]$
(which is the off-shell contribution) is even.
Making use of these additional remarks,
the collision term simplifies to
\begin{align}
{\cal C}_\ell^{{\rm v }\ell N2}
=&-[{Y_1^*}^2 Y_2^2-Y_1^2{Y_2^*}^2] \int \frac{dk^0}{2\pi}
\frac{d^4 p}{(2\pi)^4}\frac{d^4q}{(2\pi)^4}
\\\notag
{\rm tr}\Big\{&
\left[
{\rm i} S_{N2}^<(-q) {\rm i} S_\ell^> (k)
-{\rm i} S_{N2}^>(-q) {\rm i} S_\ell^< (k)
\right]
 {\rm i}\Delta_\phi^T(-p-k)
\\\notag
\times&
\big[
{\rm i}S_{N1}^<(-p) C\left[S_\ell^<(p+k+q)\right]^tC^\dagger
{\rm i} \Delta_\phi^<(-q-k)
\\\notag
-&
{\rm i}S_{N1}^>(-p) C\left[S_\ell^>(p+k+q)\right]^tC^\dagger
{\rm i} \Delta_\phi^>(-q-k)
\big]
\Big\}\,.
\end{align}
The source term is
\begin{align}
S^{{\rm v}\ell N2}
=\int\frac{d^3 p^\prime}{(2\pi)^3}
{\cal C}_\phi^{{\rm v }\ell N2}(\mathbf p^\prime)\,,
\end{align}
which becomes, when
substituting the finite-density propagators,
\begin{align}
\notag
S^{{\rm v}\ell N_2}
=&
4{\rm Im}[Y_1^2{Y_2^*}^2]
\int\frac{d^3 k^{\prime\prime}}{(2\pi)^3 2|\mathbf k^{\prime\prime}|}
\frac{d^3 p}{(2\pi)^3 2\sqrt{\mathbf p^2+M_1^2}}
\frac{d^3 p^{\prime\prime}}{(2\pi)^3 2|\mathbf p^{\prime\prime}|}
\frac{d^3 k^\prime}{(2\pi)^3 2\sqrt{{\mathbf k^\prime}^2+M_2^2}}
\frac{d^3 p^{\prime}}{(2\pi)^3 2|\mathbf p^{\prime}|}
\\
\label{S:phi}
\times&
(2\pi)^4\delta^4(p-k^{\prime\prime}-p^{\prime\prime})
(2\pi)^4\delta^4(k^\prime-k^{\prime\prime}-p^{\prime})
\frac{p^{\prime\prime}_\mu {p^\prime}^\mu}{(p^{\prime\prime}+k^\prime)^2}M_1 M_2
\\\notag
\times&
\delta f_{N1}({\mathbf p})
\left[
1-f^{\rm eq}_\ell(\mathbf p^{\prime\prime}) +f^{\rm eq}_\phi(\mathbf k^{\prime\prime})]
\right]
\times
\left[f^{\rm eq}_\ell (\mathbf p^\prime) -f^{\rm eq}_{N2}(\mathbf k^\prime)\right]\,.
\end{align}
Again, we note that the factor
$[f^{\rm eq}_\ell-f^{\rm eq}_{N2}]=-[1-f^{\rm eq}_\ell]f^{\rm eq}_{N2}+f^{\rm eq}_\ell[1-f^{\rm eq}_{N2}]$
could also be guessed from a would-be factor
$[1-f^{\rm eq}_\ell][1-f^{\rm eq}_{N2}]-f^{\rm eq}_\ell f^{\rm eq}_{N2}=1-f^{\rm eq}_\ell f^{\rm eq}_{N2}$, which would arise if $\phi\to\ell N_2$ was kinematically 
allowed,
and the replacements $-f^{\rm eq}_{N2}\leftrightarrow[1-f^{\rm eq}_{N2}]$.
We also  note that $S^{{\rm v}\ell N_2}$ is Maxwell suppressed for $M_2\gg T$, in analogy
with what is discussed $S^{{\rm v}\phi N_2}$.

\section{Examples}
\label{sec:examples}

We now present results from the numerical evaluation of the source
terms~(\ref{coll:vertex:final},\ref{S:ell},\ref{S:phi}).
In Section~\ref{subsection:effective},
we define an effective measure for the $CP$ violation from the various
cut contributions
through a benchmark out-of-equilibrium distribution for $N_1$.
Then, we evaluate this effective
$CP$ violation
as a function of $z=M_1/T$ for various values of $M_1/M_2$. We find that
in case $M_2>M_1$ the new corrections are only significant for $z\sim 1$
or smaller.
If $M_2<M_1$, the corrections from the new cuts can be relevant for larger
values
of $z$. We proceed in Section~\ref{subsection:asymmetry} with the calculation
of the lepton asymmetry by solving the kinetic Boltzmann-type
equations~(\ref{kin:eq})
with the new source terms. Our choice of the washout strength is motivated
by the wish to exhibit models where the contributions from the new cuts
are sizable, that is where relevant contributions to the final asymmetry arise
at values for $z$ around one. This is the case in the transitional regime
from weak to strong washout. We find that indeed, the new cut contributions
can have a sizable impact on the final asymmetry.  However, in case sizable
contributions to the asymmetry originate from $z\ll 1$ in our
simulations, the quantitative results need to be interpreted with care.
This is because in these regions, thermal corrections to the masses and widths of
$\phi$ and $\ell$ will become relevant.

\subsection{Effective $CP$-Violating Parameter}
\label{subsection:effective}

In order to obtain a quantitative comparison of the amount of $CP$ violation from the various source terms,
it is useful to define a benchmark form for the distribution $\delta f_{N1}(\mathbf k)$. This is necessary, since the precise form of $\delta f_{N1}(\mathbf k)$ depends on time, washout strength and initial conditions.
We follow Ref.~\cite{Beneke:2010wd} by taking for $f_{N1}(\mathbf k)$ a Fermi-Dirac
distribution with a pseudo-chemical potential $\mu_{N1}$. The deviation from
equilibrium is then obtained by expanding to linear order in $\mu_{N1}/T$,
\begin{align}
\label{delta:fN1:appr}
\delta f_{N1}(\mathbf k)
=f_{N1}^{\rm eq}(\mathbf k)\left(1-f_{N1}^{\rm eq}(\mathbf k)\right)\frac{\mu_{N1}}{T}\,.
\end{align}
When the only interactions of the neutrino $N_1$ are mediated by $Y_1$, the actual
distribution function is not exactly described by the pseudo-chemical potential, and
Eq.~(\ref{delta:fN1:appr}) should indeed only be regarded as a useful benchmark for the purpose of comparing
the various contributions to the source term.
Note however that in case there are fast elastic scatterings between the neutrinos $N_1$,
Eq.~(\ref{delta:fN1:appr}) should be a very accurate description for the actual distribution.
We substitute Eq.~(\ref{delta:fN1:appr}) into
Eqs.~(\ref{coll:vertex:final},\ref{S:ell},\ref{S:phi}) and into
$S^{\rm wf}$ and $S_{M2\gg M1}$ as given in Ref.~\cite{Beneke:2010wd}, where
$S_{M2\gg M1}$ is the source term including both, vertex and wave function terms,
evaluated in the hierarchical limit.
Out of these, we take the ratios
\begin{align}
\label{ratio:v}
(S^{{\rm v}\ell\phi}+S^{{\rm wf}})/S_{M_2\gg M_1}
\end{align}
and
\begin{align}
\label{ratio:full}
(S^{\rm v}+S^{{\rm wf}})/S_{M_2\gg M_1}
\,,
\end{align}
as functions of $z=M_1/T$ and with $S^{\rm v}$ defined in Eq.~(\ref{S:v}).
The ratio~(\ref{ratio:v}) allows for a comparison of the source term at finite density and
the standard cut through $\ell$ and $\phi$ only with the source term in the hierarchical limit. Through
the ratio~(\ref{ratio:full}), we compare the source with all cuts at finite
density with the hierarchical limit. To quantify the effect of the new cuts,
both ratios~(\ref{ratio:v}) and~(\ref{ratio:full}) are compared with one another.

Before we present the numerical results for Eqs.~(\ref{ratio:v}) and~(\ref{ratio:full}),
a few remarks on their relevance and range of validity are in order.
First, note that the relevant contributions to leptogenesis are generated when $z\sim 1$ or larger. (This is to be understood in the sense that while $z$ is
\emph{much}
smaller than one, no sizable contributions to the final asymmetry are 
generated).
This is a consequence of the fact that $N_1$ must become non-relativistic and Maxwell suppressed
before it equilibrates, because otherwise the lepton asymmetry would be completley washed out.
Therefore, the effect of the new contributions is more relevant if it extends to
larger values of $z$.

Second, we remark that
that  for small $z$, finite temperature effects should become relevant, which we do not take into account within the present work.
The most important correction is the contribution of the top quarks to
the Debye mass-square of the Higgs boson, $m^2_H=(1/4)h_t^2 T^2$, where $h_t$ is the top-quark 
Yukawa coupling. Besides, also the ${\rm SU}(2)_{\rm L}$ gauge couplings of the Higgs bosons
and the  leptons are of relevance. A full evaluation of these finite-temperature effects
in the context of leptogenesis
has not been performed yet, but it would be of great importance for obtaining quantitatively accurate 
results in models where asymmetries generated at values of $z$ that
are somewhat smaller than one are 
relevant. Moreover, the size of the top-quark Yukawa coupling at the energy scale of
leptogenesis depends due to running on the mass of the Higgs boson, which is yet unknown. For the present 
discussion, we therefore keep in mind that for the contributions that are generated at $z<1$,
we have to anticipate an inaccuracy of order one.

%\begin{figure}
%\psfrag{MATHS}{$\alpha$}
%\epsfig{file=YM2g1.1.eps,width=7cm}
%\includegraphics[width=7cm]{YM2g1.1.eps}
%\end{figure}

\begin{figure}[t!]
\begin{center}
\begin{tabular}{c}
(A): $M_2/M_1=1.1$\\%\vspace{.1cm}\\
%\includegraphics[width=7cm]{file=YM2g1.1.ps}
%\psfrag{epseff}{${\varepsilon_{\rm eff}}\big{/}{\varepsilon_{\rm eff}^{M2\gg M1}}$}
\psfrag{epseff}{$\frac{S}{S_{M2 \gg M1}}$}
\psfrag{z}{$z$}
\epsfig{file=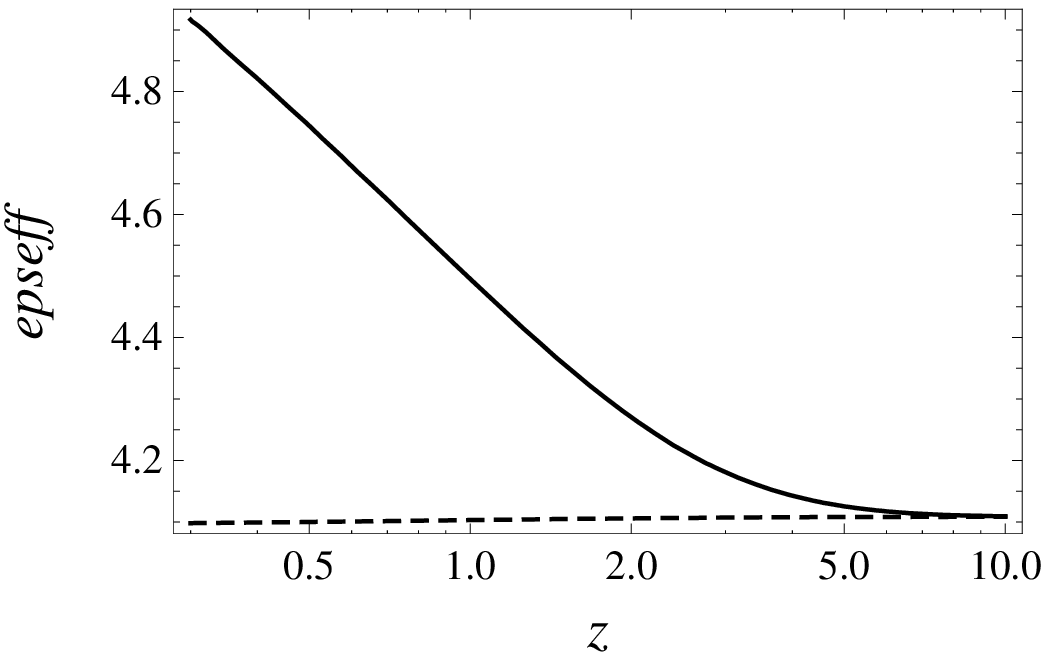,scale=.70}
\end{tabular}
\begin{tabular}{c}
(B): $M_2/M_1=2.0$\\%\vspace{.1cm}\\
%\psfrag{epseff}{${\varepsilon_{\rm eff}}\big{/}{\varepsilon_{\rm eff}^{M2\gg M1}}$}
\psfrag{epseff}{$\frac{S}{S_{M2 \gg M1}}$}
\psfrag{z}{$z$}
\epsfig{file=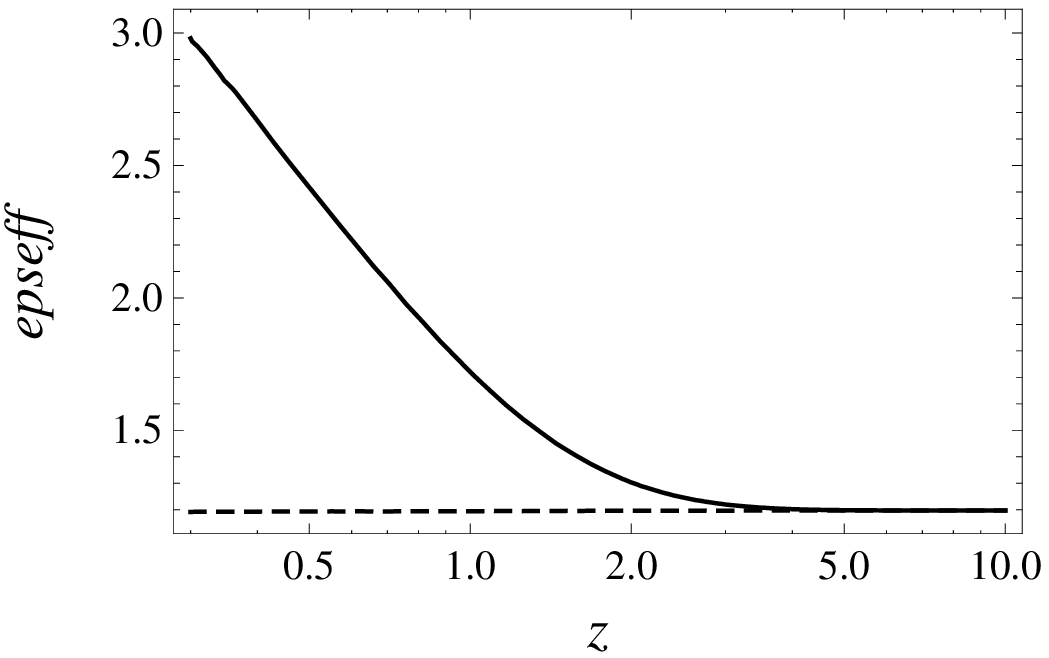,scale=.70}
\end{tabular}
\\
\begin{tabular}{c}
(C): $M_2/M_1=5.0$\\%\vspace{.1cm}\\
%\psfrag{epseff}{${\varepsilon_{\rm eff}}\big{/}{\varepsilon_{\rm eff}^{M2\gg M1}}$}
\psfrag{epseff}{$\frac{S}{S_{M2 \gg M1}}$}
\psfrag{z}{$z$}
\epsfig{file=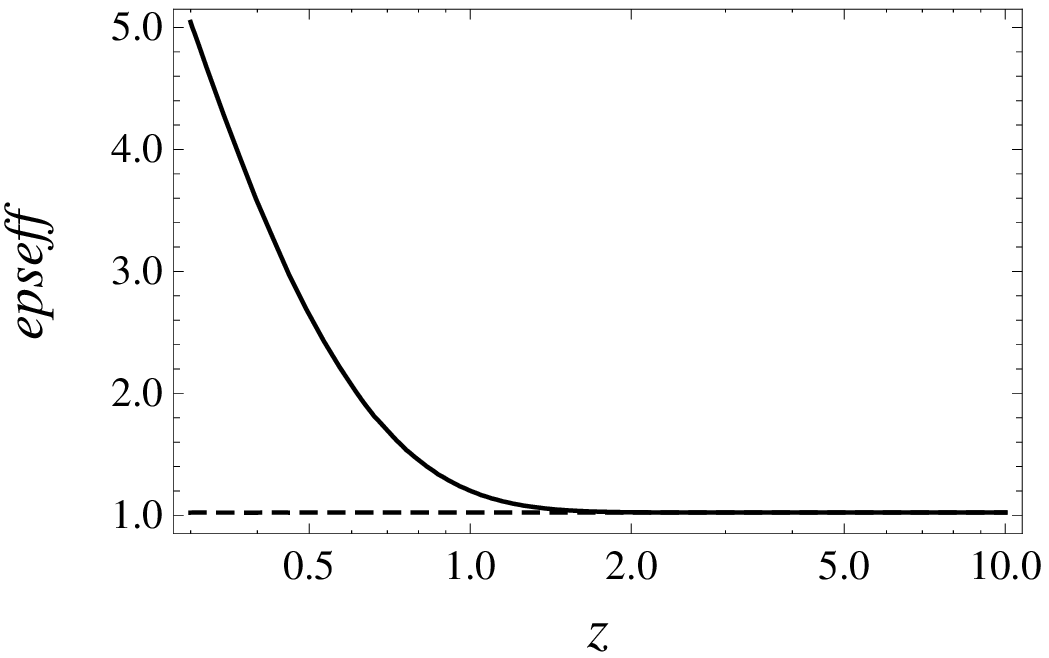,scale=.70}
\end{tabular}
\begin{tabular}{c}
(D): $M_2/M_1=0.5$\\%\vspace{.1cm}\\
%\psfrag{epseff}{${\varepsilon_{\rm eff}}\big{/}{\varepsilon_{\rm eff}^{M2\gg M1}}$}
\psfrag{epseff}{$\frac{S}{S_{M2 \gg M1}}$}
\psfrag{z}{$z$}
\epsfig{file=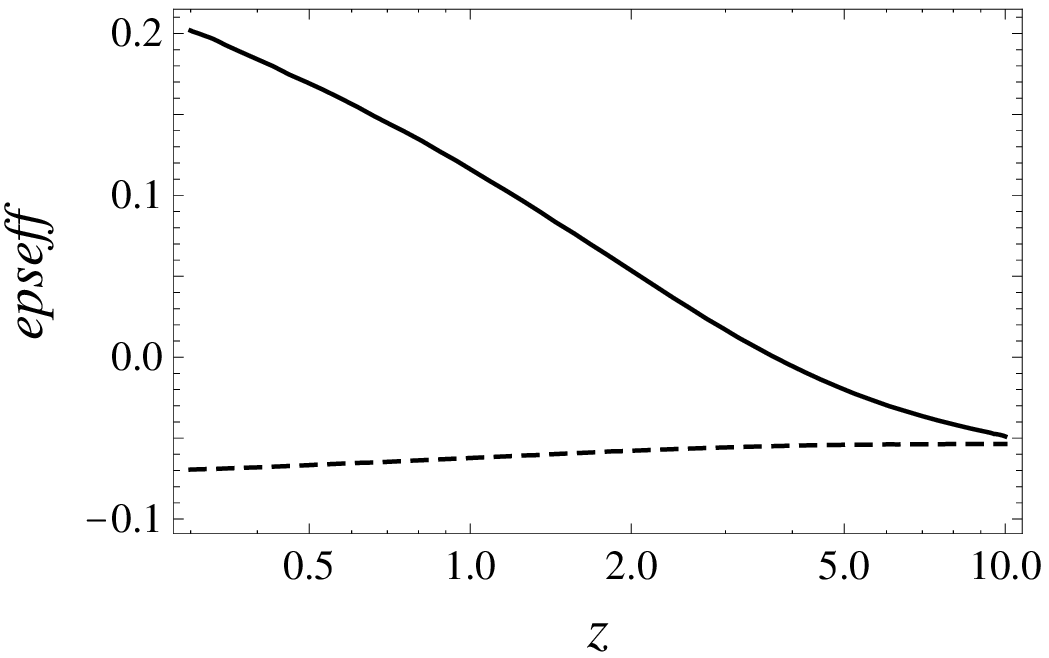,scale=.70}
\end{tabular}
\end{center}
\caption{
\label{fig:epseff}
$(S^{\rm v}+S^{{\rm wf}})/S_{M_2\gg M_1}$ (solid line) and
$(S^{{\rm v}\ell\phi}+S^{{\rm wf}})/S_{M_2\gg M_1}$ (dashed line) over $z=M_1/T$.
}
\end{figure}

In Figure~\ref{fig:epseff}, the ratios~(\ref{ratio:v}) and~(\ref{ratio:full})
are plotted for several values of $M_1/M_2$.
The main features for each of the particular values can be summarised as follows:
\begin{itemize}
\item[(A):] For $M_2/M_1=1.1$, the hierarchical limit is of course not a 
good approximation.
The total asymmetry is dominated by the resonantly enhanced contribution $S^{\rm wf}$.
As a consequence of this, $(S^{{\rm v}\ell\phi}+S^{{\rm wf}})/S_{M_2\gg M_1}$ is much larger than
one. Since the distribution of $N_2$ is only very weakly thermally suppressed when
compared to $N_1$, $S^{{\rm v}\ell\phi}+S^{{\rm wf}}$ and 
$S^{\rm v}+S^{{\rm wf}}$ agree only for comparably large values of $z$.
\item[(B):]
For $M_2/M_1=2.0$, we note that the hierarchical limit becomes a better approximation
to $S^{{\rm v}\ell\phi}+S^{{\rm wf}}$, while yet deviating by about 20\%.
Compared to case~(A), $S^{{\rm v}\ell\phi}+S^{{\rm wf}}$ and 
$S^{\rm v}+S^{{\rm wf}}$ start to agree for smaller values of $z$, as the distribution
of $N_2$ now suffers from stronger thermal suppression. We note that the very substantial
deviation of $S^{{\rm v}\ell\phi}+S^{{\rm wf}}$ from
$S^{\rm v}+S^{{\rm wf}}$ for very small values of $z$ needs to be interpreted
with care in the light of the thermal corrections mentioned above.
\item[(C):]
For $M_2/M_1=5.0$, the results obtained from the hierarchical limit $S_{M2\gg M_1}$ and
the source $S^{{\rm v}\ell\phi}+S^{{\rm wf}}$ with the standard cuts only
are now in good agreement, as anticipated. This serves also as a
consistency check for our numerical evaluation of $S^{\rm v}+S^{{\rm wf}}$
in the form of a multi-dimensional integral as compared to
$S_{M2\gg M1}$ through the one-dimensional integral given in Ref.~\cite{Beneke:2010wd}.
The new cut contributions become suppressed
for even smaller values of $z$ along with the stronger Maxwell suppression of $N_2$.
The comment regarding Scenario~(B), that the deviations that originate for values $z\ll1$ need
to be interpreted with care due to thermal corrections, is even more relevant for the 
present case.
\item[(D):]
In this Scenario, $M_2/M_1=0.5$. Since $N_2$ is now lighter than $N_1$,
the coupling $Y_2$
must be chosen sufficiently small, such that $N_2$ does not equilibrate and wash out the
lepton asymmetry before it becomes non-relativistic and Maxwell suppressed.
In this situation, for
the whole relevant range of $z$, neither
the hierarchical limit nor the sources $S^{{\rm v}{\ell\phi}}+S^{{\rm wf}}$ are an accurate approximation
to the full result $S^{\rm v}+S^{{\rm wf}}$.
\end{itemize}

\subsection{Lepton Asymmetries}
\label{subsection:asymmetry}

From the comparison of the effective amount of $CP$ violation in presence
of the new cuts with the case when the new cuts are neglected, we see
that when $M_2>M_1$, the deviations are only sizable for $z$ is of order a few
or smaller. For larger
values of $z$, the distribution of the $N_2$ is strongly Maxwell suppressed,
and the new cut contributions become irrelevant. In order to exhibit
the effect of the new cuts, we therefore choose the parameters $M_1$ and
$Y_1$ such that leptogenesis takes place in the transitional regime from
weak to strong washout. In this situation, sizable contributions to the
final lepton asymmetry are generated for $z\sim 1$, such that the
effects of the new cuts becomes relevant.

We choose for $N_{1,2}$ thermal initial conditions. In the early Universe,
these may be established through interactions via heavy gauge bosons that
freeze out at times before leptogenesis takes place. We assume
that $N_2$ is maintained in equilibrium due to a Yukawa coupling with an 
additional lepton flavour, within which no asymmetry is produced.
The coupling $Y_2$ is chosen smaller than $Y_1$, such that washout effects from
inverse decays of $N_2$ are negligible. (Explicitly, for the scenarios with $M_2>M_1$,
the largest error from washout through $Y_2$ occurs for $M_2/M_1=1.1$. Since
the thermal suppression of $N_2$ compared to $N_1$ is very small in this case
and we have chosen $Y_2/Y_1=1/2$, we expect that the washout is underestimated
by 20\%. The accuracy improves for larger ratios of $M_2>M_1$ due to the
thermal suppression of $N_2$ and its irrelevance for washout).
While this appears as a somewhat special setup, we note that even when $Y_2$ is
large, the vector of couplings of $N_2$ to the various left-handed lepton flavours
defines a particular linear combination $\ell_2$
of leptons that are washed out through inverse
decays of $N_2$. This linear combination can in general be linearly independent
of the linear combination $\ell$
within which the lepton asymmetry through decays and inverse
decays of $N_1$ is produced. The contribution to the asymmetry in $\ell$
that is orthogonal to $\ell_2$ is
then unaffected by the washout through $N_2$~\cite{Engelhard:2006yg}. (Note
the different assignment of the heavy neutrinos to the indices $1,2$ in that work).
Therefore, the qualitative features of our particular setup should be relevant
for parametrically more generic models of leptogenesis.
However, it would still be interesting to
study the effect of different initial conditions and the possibility of $N_2$ being
out-of-equilibrium within future work.

As it is described in detail in Ref.~\cite{Beneke:2010wd}, we obtain the
numerical results as follows: First, we solve the evolution equations
for $f_{N1}(\mathbf k)$ as a function of $z$. These, we feed into
the washout term $W$ and the source term $S$ for the leptons, in order to
solve for the lepton asymmetry in Eq.~(\ref{kin:eq}). The expansion of the Universe is taken
into account when inserting the scale factor according to
$M_{1,2}\to a M_{1,2}$.

\begin{figure}[t]
\begin{center}
\begin{tabular}{c}
\hspace{2.cm}(A): $M_2/M_1=1.1$, $Y_2=10^{-2}$\vspace{.1cm}\\
\psfrag{Yell}{$Y_\ell$}
\psfrag{z}{$z$}
\epsfig{file=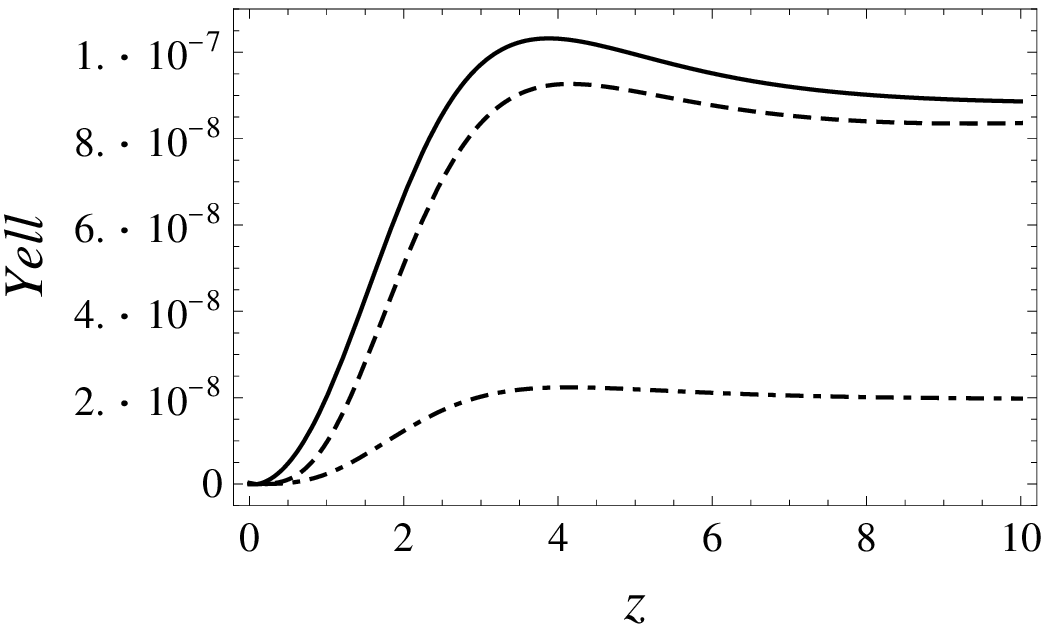,scale=.7}
\end{tabular}
\begin{tabular}{c}
\hspace{2.cm}(B): $M_2/M_1=2.0$, $Y_2=10^{-2}$\vspace{.1cm}\\
\psfrag{Yell}{$Y_\ell$}
\psfrag{z}{$z$}
\epsfig{file=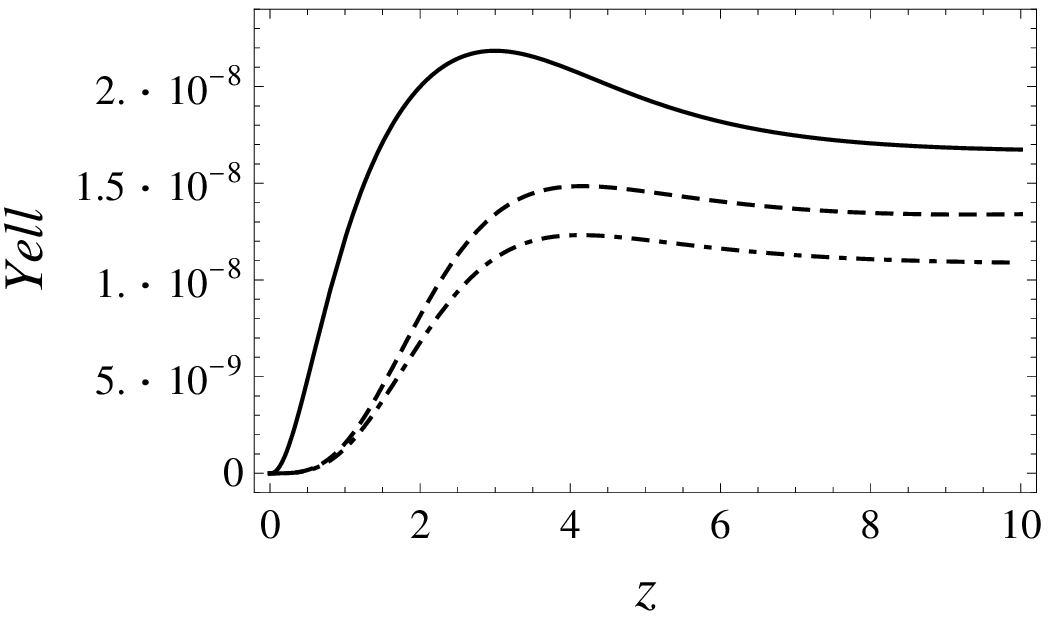,scale=.7}
\end{tabular}
\\
\begin{tabular}{c}
\hspace{2.cm}(C): $M_2/M_1=5.0$, $Y_2=10^{-2}$\vspace{.1cm}\\
\psfrag{Yell}{$Y_\ell$}
\psfrag{z}{$z$}
\epsfig{file=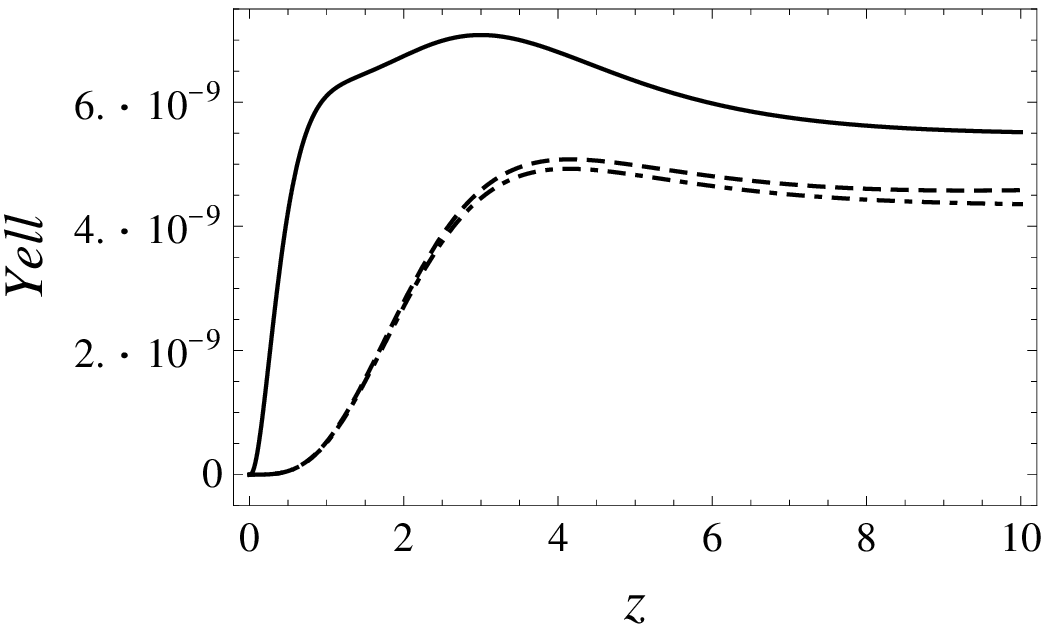,scale=.7}
\end{tabular}
\begin{tabular}{c}
\hspace{2.cm}(D): $M_2/M_1\!\!=\!0.5$, $Y_2\!\!=\!5\!\times\!\!10^{-3}$\vspace{.1cm}\\
\psfrag{Yell}{$Y_\ell$}
\psfrag{z}{$z$}
\epsfig{file=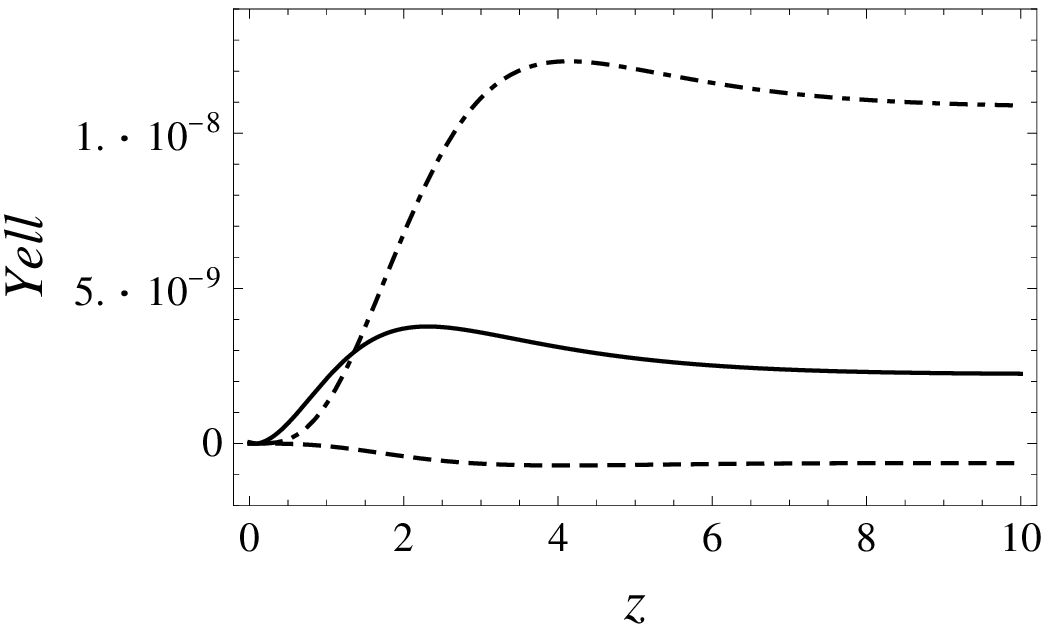,scale=.7}
\end{tabular}
\end{center}
\caption{
\label{figure:asymmetries}
Evolution of the lepton asymmetry $Y_\ell$ over $z=M_1/T$. The choice of
parameters is $M_1=10^{13}{\rm GeV}$, $Y_1=2{\rm i}\times 10^{-2}$.
Solid: full result; dashed: result with contribution from cut through
$\{\ell\,,\phi\}$ only; dot dashed: hierarchical limit $M_2\gg M_1$.
}
\end{figure}

The results for the lepton-number to entropy ratio $Y_\ell$ as
defined in Ref.~\cite{Beneke:2010wd}
are presented in Figure~\ref{figure:asymmetries}. Note that
it is instructive to compare the particular panels with those in
Figure~\ref{fig:epseff}. Again, we summarise some features for each
of the particular values of $M_2/M_1$:
\begin{itemize}
\item[(A):] For $M_2/M_1=1.1$, the hierarchical limit clearly underestimates
the full result. This is anticipated, because the wave-function contribution
to the full result is resonantly enhanced for $M_1\approx M_2$. The new
cuts give rise to relative corrections at the 10\% level. Note however
that the absolute correction is larger when compared to scenarios (B) and (C).
The relative correction is marginalised due to the resonant enhancement
of $S^{\rm wf}$.
\item[(B):] For $M_2/M_1=2.0$, there are sizable deviations of the
full result from the result with the cut through $\{\ell,\phi\}$ only, that
arise in the region $z\sim 1$.
\item[(C):] For $M_2/M_1=5.0$, the new contributions are of importance
for smaller values of $z$ when compared to scenarios (A) and (B). In the
light of the thermal corrections that we anticipate to be important for
small values of $z$, the quantitative result needs to be interpreted
with care. The good agreement between the result from the standard cut
through $\{\ell,\phi\}$ only and the hierarchical limit serves again
as a consistency check for our numerical evaluations.
\item[(D):] For $M_2/M_1=0.5$, we choose a smaller value for $Y_2$,
motivated by the requirement that the $N_2$ must not wash out the lepton
asymmetry. The full result and
the result with the cut through $\{\ell\,,\;\phi\}$ only receive different
contributions even for values of $z$ of the order of a few,
{\it cf.} Figure~\ref{fig:epseff}~(D). We expect
therefore only a small contamination from theoretical uncertainties due to
thermal corrections. Note that in this scenario most
of the lepton asymmetry of the Universe originates from the new cuts.
\end{itemize}

\section{Flavoured Leptogenesis}
\label{sec:flavour}

\begin{figure}[t!]
\begin{center}
%\begin{tabular}{c}
\epsfig{file=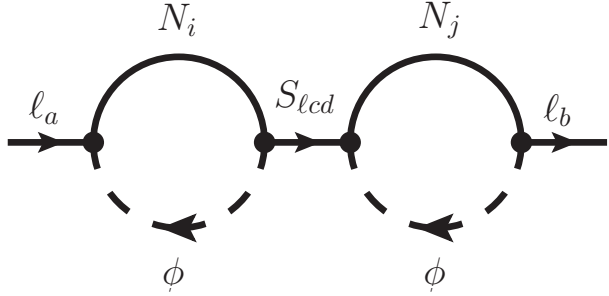,width=9.cm}
%\hskip.5cm
\end{center}
\caption{
\label{fig:flavourwf}
Wave function correction that contributes to the lepton asymmetry in flavoured leptogenesis.
}
\end{figure}

When we insert a loop of $\phi$ and $N_2$ as a wave-function correction
into the propagator $S_\ell$ in the vacuum, no branch cut term due to on-shell
$\phi$ and $N_2$ arises for kinematic reasons. Again, this holds no longer
true in a finite temperature background. Therefore, $CP$-violating source terms
can arise from the diagram in Figure~\ref{fig:flavourwf}.
Since there is no lepton number violation, this does not yield a contribution
to the asymmetry in models of unflavoured leptogenesis.
However, this diagram may appear as a source in flavoured scenarios.
In this Section, we give a rough estimate of this
contribution, leading to the conclusion that it is generically negligible.

Within the CTP-formalism,
the form of the Wightman type self-energy, that is given diagrammatically
in Figure~\ref{fig:flavourwf}, reads
\begin{align}
{\rm i}{\Sigma\!\!\!\!/}^{\,{\rm wf}\ell >}_{\ell ab}(k)=&
-Y_{ia}^*Y_{ic}Y_{jd}^*Y_{jb}
\int\frac{d^4 p}{(2\pi)^4}\frac{d^4 q}{(2\pi)^4}
\\\notag
\Big\{
&
{\rm i} S_{Ni}^>(p)
{\rm i} \Delta_\phi^<(p-k)
{\rm i} S^T_{\ell cd}(k)
{\rm i} S_{Nj}^T(q)
{\rm i} \Delta_\phi^T(q-k)
\\\notag
-&
{\rm i} S_{Ni}^>(p)
{\rm i} \Delta_\phi^<(p-k)
{\rm i} S^<_{\ell cd}(k)
{\rm i} S_{Nj}^>(q)
{\rm i} \Delta_\phi^<(q-k)
\\\notag
-&
{\rm i} S_{Ni}^{\bar T}(p)
{\rm i} \Delta_\phi^{\bar T}(p-k)
{\rm i} S^>_{\ell cd}(k)
{\rm i} S_{Nj}^T(q)
{\rm i} \Delta_\phi^T(q-k)
\\\notag
+&
{\rm i} S_{Ni}^{\bar T}(p)
{\rm i} \Delta_\phi^{\bar T}(p-k)
{\rm i} S^{\bar T}_{\ell cd}(k)
{\rm i} S_{Nj}^>(q)
{\rm i} \Delta_\phi^<(q-k)
\Big\}\,.
\end{align}
Compared to the unflavoured scenario, we have promoted the Yukawa couplings
of the right-handed neutrinos $N_i$
to a matrix $Y_{ia}$, where the first index refers to the right-handed neutrino
and the second index to the left-handed lepton flavour. For the definitions
of the model Lagrangian and the lepton propagator $S_{\ell}$, we refer
to Ref.~\cite{Beneke:2010dz}.

Next, we insert ${\Sigma\!\!\!\!/}^{\,{\rm wf}\ell >}_{\ell ab}$
into the collision term~(\ref{collision:KB}).
This again simplifies when substituting
equilibrium propagators for $N_2$, $\ell$ and $\phi$ and exploiting
KMS relations. We furthermore choose to work in the flavour basis where
the matrix of
Standard Model lepton Yukawa couplings $h_{ab}$ is diagonal. This is
advantageous since off-diagonal components of
$S_{\ell ab}$ are damped away quickly in this basis, provided the interactions
mediated by $h_{ab}$ are fast compared to the Hubble
rate~\cite{Endoh:2003mz,Pilaftsis:2005rv,Abada:2006fw,Nardi:2006fx},
{\it cf.} Ref.~\cite{Beneke:2010dz} for a description within
the CTP approach and for a numerical study of
this effect. Note that this also implies that
the off-diagonal components of the equilibrium propagator for
$\ell$ are vanishing. We eventually obtain for the collision term
\begin{align}
\label{C:flav:wf}
{{\cal C}_{\ell aa}^{{\rm wf}\ell}}(\mathbf k)
=&-[Y_{1a}^*Y_{1c}Y_{2c}^*Y_{2a}-Y_{2a}^*Y_{2c}Y_{1c}^*Y_{1a}]
\int\frac{dk^0}{2\pi}\frac{d^4 p}{(2\pi)^4}\frac{d^4 q}{(2\pi)^4}
\\\notag
\frac 12{\rm tr}
\Big\{
&\left(
{\rm i}S_{\ell aa}^<(k)
{\rm i} S_{Ni}^>(p)
{\rm i} \Delta_\phi^<(p-k)
-
{\rm i}S_{\ell aa}^>(k)
{\rm i} S_{Ni}^<(p)
{\rm i} \Delta_\phi^>(p-k)
\right)
{\rm i}{\rm Re}\left[\frac{1}{k^2+{\rm i\varepsilon}}\right]
%{\rm i} S^T_{\ell cc}(k)
\\\notag
\times&
\left(
{\rm i} S_{Nj}^<(q)
{\rm i} \Delta_\phi^>(q-k)
-
{\rm i} S_{Nj}^>(q)
{\rm i} \Delta_\phi^<(q-k)
\right)
\Big\}\,,
\end{align}
where $\varepsilon$ is infinitesimal.
In the zero-width limit, this term contains a divergence from the factor
$1/(k^2+{\rm i}\varepsilon)$
that originates from the propagator
$S^T_{\ell}(k)$, since $S_{\ell aa}^<(k)$ is proportional to
$\delta(k^2)$.
At finite temperature, this is regulated through the replacements
\begin{align}
\label{subst:ST}
\frac1{k^2+{\rm i}\varepsilon}\to\frac1{k^2-m_{\ell c}^2+{\rm i}k^0\Gamma_{\ell}}
\end{align}
and
\begin{align}
\label{subst:delta}
\delta(k^2)\to
\frac{\rm i}{2\pi}\left[
\frac1{k^2-m_{\ell a}^2+{\rm i}k^0\Gamma_{\ell}}-
\frac1{k^2-m_{\ell a}^2-{\rm i}k^0\Gamma_{\ell}}
\right]\,.
\end{align}
Here, the thermal masses of the leptons are
$m^2_{\ell a}= h_{aa}^2 \varsigma^{{\rm fl},h}(k) + g^2 \varsigma^{{\rm bl},g}(k)$,
 where $g$ is the ${\rm SU}(2)_{\rm L}$ gauge coupling.
The functions $\varsigma^{{\rm fl},h}(k)$ and $\varsigma^{{\rm bl},g}(k)$ are of
order $T^2$ for $k$ of order $T$. They account for thermal mass corrections
from flavour-sensitive and flavour blind interactions, and are discussed in more
detail in Ref.~\cite{Beneke:2010dz}. Likewise $\Gamma_\ell=g^2 g^{{\rm bl}}(k)$
is the finite-temperature width of the leptons, where
$ g^{{\rm bl}}(k)$ is of order $T$ when $k$ of order $T$.
When leptogenesis occurs at temperatures of roughly below $10^{11}$ to
$10^{12}{\rm GeV}$ but above $10^{8}$ to
$10^{9}{\rm GeV}$, the
$\tau$-lepton Yukawa coupling $h_\tau$
is in equilibrium, while the electron $e$ and muon $\mu$
couplings are yet out-of-equilibrium. In this situation, flavoured leptogenesis
distinguishes effectively between two flavours, where flavour 1
is identified with $\tau$ and the coupling $h_{11}=h_\tau$
and flavour 2 with a linear combination
of $e$ and $\mu$ and with negligible Yukawa coupling $h_{22}\approx0$.
% For a consistent expansion also $\Delta_\phi$ should be evaluated including thermal mass and finite 
% width corrections. Since we cannot make use of on-shell $\delta$ functions in the present situation,
% the integral~(\ref{C:flav:wf}) is harder to evaluate than
The integral~(\ref{C:flav:wf}) could possibly again be evaluated numerically. However, we
can estimate that
from the factors~(\ref{subst:ST}) and~(\ref{subst:delta}), the $k^0$-integration yields
a factor
\begin{align}
\frac{m_1^2-m_2^2}{T^3 \Gamma_{\rm \ell}^2}\sim
\frac{1}{T^3}\frac{h_\tau^2}{g^4}\,,
\end{align}
for $|\mathbf k|\sim T$.  The latter estimate follows from
above assumption that only $h_\tau$ appears as a relevant Standard Model Yukawa
coupling. Above suppression factor is to be
compared to $1/T^3$ for a leptogenesis scenario with $M_1\sim M_2$ but
no pronounced resonant enhancement.
Since $h_\tau^2/g^4\ll 1$, we conclude that this contribution to
flavoured leptogenesis is suppressed due to the large width of
the leptons $\ell$ at finite temperature. In other words, the contribution
to the $CP$-asymmetry is rendered ineffective because the separation
between the resonances of the lepton quasi-particles is well within their overlap
due to the finite widths.

\section{Conclusions}
\label{sec:conclusions}

In this paper, we have presented in the form of Eqs.~(\ref{S:ell}) and~(\ref{S:phi})
the first results for source terms that contribute to
the lepton asymmetry in a finite-density background, but that are absent in the vacuum.
In order for these to
be relevant, we have seen that $M_2$ should not be much larger than $T$ at the time of
leptogenesis. The main features of the numerical evaluations are
easily understood. First, $M_2$ should not be much larger than $M_1$ for the new
effects to be sizable at larger values of $z$, {\it cf.} Figure~\ref{fig:epseff}.
We reemphasise that the quantitative results for the
scenario in Figure~\ref{figure:asymmetries}~(C) should be considered
with great care since a substantial amount of the deviations incurred by the
new effects is generated at very small values of $z$, where we expect thermal
corrections to become relevant. Second, the effect is most pronounced
in scenarios where a sizable amount of the asymmetry is produced for comparably
small values of $z$, as it is the case for the transitional regime between weak and
strong washout. As a loophole, we find that the largest effects arise within a somewhat unconventional scenario
with $M_2<M_1$, since in this situation the new cuts are important
throughout the time of the out-of-equilibrium decays of $N_1$. Therefore,
relevant contributions from the new cuts also result for larger values of $z$.
However,
in regard of Ref.~\cite{Engelhard:2006yg}, where it is shown that decay asymmetries
from the heavier singlet neutrinos generically survive subsequent washout,
such a scenario may not only appear as a mere loophole.
Provided the reheat temperature is large enough to produce the heavier singlet neutrino
in the early Universe, the present work implies that a large contribution to
the baryon asymmetry of the Universe generically results from the new cuts.

We mention that while this work has been in preparation, Ref.~\cite{Fong:2010bh}
appeared, where non-standard cuts are discussed as well. Since finite density effects
are neglected, it is found there that
the new cuts are only allowed within certain scattering processes.
More precisely, in order to satisfy
the kinematic thresholds for the presence of cut contributions in the vacuum,
diagrams with additional radiation of Standard Model particles are considered.
Note that if the resulting amplitudes were substituted as source terms into the
kinetic equations for the lepton asymmetry, they would be subject to the
same Maxwell suppression that occurs also in the present case for $M_2\gg T$.
However, compared to the new source terms that we have derived in the present work,
the terms discussed in Ref.~\cite{Fong:2010bh} are subject to
further suppression because of the insertion of additional coupling constants.

On the ground of the present results, it would be interesting to address the
following points in the future:
\begin{itemize}
\item
The results~(\ref{S:ell}) and~(\ref{S:phi}) should be generalised to include effects
from deviations of the distribution $f_{N2}$ from equilibrium, which we expect to be
non-zero in generic scenarios of leptogenesis.
\item
Systematic investigations of the parameter space for scenarios where the new
cuts are relevant would be desirable. We expect the new cut contributions
to be of crucial importance for phenomenological studies where it is assumed
that $M_2<M_1$, such as Refs.~\cite{DiBari:2005st,Engelhard:2006yg,Bertuzzo:2010et},
where $N_2$ may or may
not be in equilibrium.
(Note the different definitions of $N_{1,2}$ in these works). Note that in case
$M_2\ll M_1$, thermal corrections may become important again, since the thermal masses
of $\phi$ and $\ell$ at the time of the decay of $N_1$ may exceed $M_2$. We expect the
new cuts still to be important in such a situation, but the interpretations
of the particular cut particles in terms of absorption processes may change
to emissions and vice versa.
\item
In the present work, we have restricted ourselves to compute the lepton asymmetries
for thermal initial conditions for $N_{1,2}$. This is in part motivated by the fact that
for a vanishing initial density of $N_1$, the final asymmetry is a remainder
of an incomplete cancellation of a contribution that is created initially
at small $z$, when $N_1$ is 
underabundant, and an opposite one through later decays when $z$ is larger and $N_1$
overabundant. The fact that the cancellation is incomplete is because the opposite
asymmetries are affected differently by the washout, since they occur at different
temperatures. In order to obtain a quantitatively reliable result for the
remaining asymmetry, a rather accurate prediction of the asymmetry that is created
through inverse decays at small $z$ is necessary. Due to the uncertainties because
of thermal corrections, that we have emphasised in this paper, such an accurate
prediction is presently not available. The situation somewhat improves for the
thermal initial conditions that we consider within the present work, because $N_1$
is always overabundant and the produced asymmetry is therefore of the same sign
for all values of $z$.
\end{itemize}

In regard of these points, we briefly comment on possible technical improvements
that may prove very useful in order to increase the accuracy of the predictions
for leptogenesis from the new cuts as well as from the standard cuts. It would be
particularly interesting, if the following issues were addressed:
\begin{itemize}
\item
As it should be clear from the discussions within this paper, the uncertainties
due to thermal corrections for small $z$ are problematic for the predictivity
of the new cut contributions as well as more generically for leptogenesis in
the weak washout regime, in particular for vanishing initial conditions for $N_1$.
In order to resolve this issue, a calculation of the rates
$N_1\leftrightarrow\ell\phi$ for temperatures that are of order $M_1$ or larger
is necessary. Both, the tree-level rates as well as the $CP$-violating loop effects
need to be calculated. For that purpose, in particular the thermal masses and
finite widths of $\ell$ and $\phi$ should be taken into account. First work into
this direction has been reported in Ref.~\cite{Kiessig:2010pr}.
\item
In Ref.~\cite{Anisimov:2010aq}, sizable effects from the finite width of $N_1$
in the $CP$-violating source term have been reported. It needs to be verified,
whether the initial conditions chosen in that work are applicable to the situation
in the early Universe. Furthermore, the finite widths of $\ell$ and $\phi$, which are 
much larger than for $N_1$ and have been neglected so far, need to be taken into
account in a future calculation.
\end{itemize}

Considering these comments and a number of related papers, the present
work may be regarded as a contribution to present efforts to improve
the theoretical description of leptogensis, to increase the
accuracy of quantitative predictions and to extend their range of applicability.

\subsubsection*{Acknowledgements}

\noindent
The author would like to thank Martin Beneke and Pedro Schwaller for useful comments
on the manuscript.
This work is supported by the Gottfried Wilhelm Leibniz programme 
of the Deutsche Forschungsgemeinschaft.


\begin{thebibliography}{99}

%\cite{Fukugita:1986hr}
\bibitem{Fukugita:1986hr}
  M.~Fukugita and T.~Yanagida,
  ``Baryogenesis Without Grand Unification,''
  Phys.\ Lett.\  B {\bf 174}, 45 (1986).
  %%CITATION = PHLTA,B174,45;%%


%\cite{Davidson:2002qv}
\bibitem{Davidson:2002qv}
  S.~Davidson and A.~Ibarra,
  ``A lower bound on the right-handed neutrino mass from leptogenesis,''
  Phys.\ Lett.\  B {\bf 535} (2002) 25
  [arXiv:hep-ph/0202239].
  %%CITATION = PHLTA,B535,25;%%

%\cite{Buchmuller:2002rq}
\bibitem{Buchmuller:2002rq}
  W.~Buchm\"uller, P.~Di Bari and M.~Pl\"umacher,
  ``Cosmic microwave background, matter-antimatter asymmetry and neutrino
  masses,''
  Nucl.\ Phys.\  B {\bf 643} (2002) 367
  [Erratum-ibid.\  B {\bf 793} (2008) 362]
  [arXiv:hep-ph/0205349].
  %%CITATION = NUPHA,B643,367;%%

%\cite{Buchmuller:2003gz}
\bibitem{Buchmuller:2003gz}
  W.~Buchm\"uller, P.~Di Bari and M.~Pl\"umacher,
  ``The neutrino mass window for baryogenesis,''
  Nucl.\ Phys.\  B {\bf 665} (2003) 445
  [arXiv:hep-ph/0302092].
  %%CITATION = NUPHA,B665,445;%%

%\cite{Beneke:2010wd}
\bibitem{Beneke:2010wd}
  M.~Beneke, B.~Garbrecht, M.~Herranen and P.~Schwaller,
  ``Finite Number Density Corrections to Leptogenesis,''
  Nucl.\ Phys.\  B {\bf 838} (2010) 1
  [arXiv:1002.1326 [hep-ph]].
  %%CITATION = NUPHA,B838,1;%%

%\cite{Flanz:1994yx}
\bibitem{Flanz:1994yx}
  M.~Flanz, E.~A.~Paschos and U.~Sarkar,
  ``Baryogenesis from a lepton asymmetric universe,''
  Phys.\ Lett.\  B {\bf 345} (1995) 248
  [Erratum-ibid.\  B {\bf 382} (1996) 447]
  [arXiv:hep-ph/9411366].
  %%CITATION = PHLTA,B345,248;%%

%\cite{Flanz:1996fb}
\bibitem{Flanz:1996fb}
  M.~Flanz, E.~A.~Paschos, U.~Sarkar and J.~Weiss,
  ``Baryogenesis through mixing of heavy Majorana neutrinos,''
  Phys.\ Lett.\  B {\bf 389} (1996) 693
  [arXiv:hep-ph/9607310].
  %%CITATION = PHLTA,B389,693;%%


%\cite{Pilaftsis:1997jf}
\bibitem{Pilaftsis:1997jf}
  A.~Pilaftsis,
  ``CP violation and baryogenesis due to heavy Majorana neutrinos,''
  Phys.\ Rev.\  D {\bf 56} (1997) 5431
  [arXiv:hep-ph/9707235].
  %%CITATION = PHRVA,D56,5431;%%

%\cite{Covi:1996wh}
\bibitem{Covi:1996wh}
  L.~Covi, E.~Roulet and F.~Vissani,
  ``CP violating decays in leptogenesis scenarios,''
  Phys.\ Lett.\  B {\bf 384} (1996) 169
  [arXiv:hep-ph/9605319].
  %%CITATION = PHLTA,B384,169;%%

%\cite{Pilaftsis:2003gt}
\bibitem{Pilaftsis:2003gt}
  A.~Pilaftsis and T.~E.~J.~Underwood,
  ``Resonant leptogenesis,''
  Nucl.\ Phys.\  B {\bf 692} (2004) 303
  [arXiv:hep-ph/0309342].
  %%CITATION = NUPHA,B692,303;%%

%\cite{Khlopov:1984pf}
\bibitem{Khlopov:1984pf}
  M.~Y.~Khlopov and A.~D.~Linde,
  ``Is It Easy To Save The Gravitino?,''
  Phys.\ Lett.\  B {\bf 138} (1984) 265.
  %%CITATION = PHLTA,B138,265;%%

%\cite{Ellis:1984eq}
\bibitem{Ellis:1984eq}
  J.~R.~Ellis, J.~E.~Kim and D.~V.~Nanopoulos,
  ``Cosmological Gravitino Regeneration And Decay,''
  Phys.\ Lett.\  B {\bf 145}, 181 (1984).
  %%CITATION = PHLTA,B145,181;%%

%\cite{Ellis:1984er}
\bibitem{Ellis:1984er}
  J.~R.~Ellis, D.~V.~Nanopoulos and S.~Sarkar,
  ``The Cosmology Of Decaying Gravitinos,''
  Nucl.\ Phys.\  B {\bf 259}, 175 (1985).
  %%CITATION = NUPHA,B259,175;%%

%\cite{Pilaftsis:2005rv}
\bibitem{Pilaftsis:2005rv}
  A.~Pilaftsis and T.~E.~J.~Underwood,
  ``Electroweak-scale resonant leptogenesis,''
  Phys.\ Rev.\  D {\bf 72} (2005) 113001
  [arXiv:hep-ph/0506107].
  %%CITATION = PHRVA,D72,113001;%%

%\cite{Blanchet:2009bu}
\bibitem{Blanchet:2009bu}
  S.~Blanchet, Z.~Chacko, S.~S.~Granor and R.~N.~Mohapatra,
  ``Probing Resonant Leptogenesis at the LHC,''
  arXiv:0904.2174 [hep-ph].
  %%CITATION = ARXIV:0904.2174;%%

%\cite{Blanchet:2009kk}
\bibitem{Blanchet:2009kk}
  S.~Blanchet, T.~Hambye and F.~X.~Josse-Michaux,
  ``Reconciling leptogenesis with observable $\mu \to e \gamma$ rates,''
  JHEP {\bf 1004} (2010) 023
  [arXiv:0912.3153 [hep-ph]].
  %%CITATION = JHEPA,1004,023;%%

%\cite{Giudice:2003jh}
\bibitem{Giudice:2003jh}
  G.~F.~Giudice, A.~Notari, M.~Raidal, A.~Riotto and A.~Strumia,
  ``Towards a complete theory of thermal leptogenesis in the SM and MSSM,''
  Nucl.\ Phys.\  B {\bf 685} (2004) 89
  [arXiv:hep-ph/0310123].
  %%CITATION = NUPHA,B685,89;%%

%\cite{Schwinger:1960qe}
\bibitem{Schwinger:1960qe}
  J.~S.~Schwinger,
  ``Brownian motion of a quantum oscillator,''
  J.\ Math.\ Phys.\  {\bf 2} (1961) 407.
  %%CITATION = JMAPA,2,407;%%

%\cite{Keldysh:1964ud}
\bibitem{Keldysh:1964ud}
  L.~V.~Keldysh,
  ``Diagram technique for nonequilibrium processes,''
  Zh.\ Eksp.\ Teor.\ Fiz.\  {\bf 47} (1964) 1515
  [Sov.\ Phys.\ JETP {\bf 20
} (1965) 1018].
  %%CITATION = SPHJA,20,1018;%%

%\cite{Buchmuller:2000nd}
\bibitem{Buchmuller:2000nd}
  W.~Buchm\"uller and S.~Fredenhagen,
  ``Quantum mechanics of baryogenesis,''
  Phys.\ Lett.\  B {\bf 483} (2000) 217
  [arXiv:hep-ph/0004145].
  %%CITATION = PHLTA,B483,217;%%

%\cite{De Simone:2007rw}
\bibitem{De Simone:2007rw}
  A.~De Simone and A.~Riotto,
  ``Quantum Boltzmann Equations and Leptogenesis,''
  JCAP {\bf 0708} (2007) 002
  [arXiv:hep-ph/0703175].
  %%CITATION = JCAPA,0708,002;%%

%\cite{Garny:2009rv}
\bibitem{Garny:2009rv}
  M.~Garny, A.~Hohenegger, A.~Kartavtsev and M.~Lindner,
  ``Systematic approach to leptogenesis in nonequilibrium QFT: vertex
  contribution to the CP-violating parameter,''
  Phys.\ Rev.\  D {\bf 80} (2009) 125027
  [arXiv:0909.1559 [hep-ph]].
  %%CITATION = PHRVA,D80,125027;%%

%\cite{Garny:2009qn}
\bibitem{Garny:2009qn}
  M.~Garny, A.~Hohenegger, A.~Kartavtsev and M.~Lindner,
  ``Systematic approach to leptogenesis in nonequilibrium QFT: self-energy
  contribution to the CP-violating parameter,''
  Phys.\ Rev.\  D {\bf 81}, 085027 (2010)
  [arXiv:0911.4122 [hep-ph]].
  %%CITATION = PHRVA,D81,085027;%%

%\cite{Garny:2010nj}
\bibitem{Garny:2010nj}
  M.~Garny, A.~Hohenegger and A.~Kartavtsev,
  ``Medium corrections to the CP-violating parameter in leptogenesis,''
  Phys.\ Rev.\  D {\bf 81} (2010) 085028
  [arXiv:1002.0331 [hep-ph]].
  %%CITATION = PHRVA,D81,085028;%%



%\cite{Anisimov:2010aq}
\bibitem{Anisimov:2010aq}
  A.~Anisimov, W.~Buchm\"uller, M.~Drewes and S.~Mendizabal,
  ``Leptogenesis from Quantum Interference in a Thermal Bath,''
  Phys.\ Rev.\ Lett.\  {\bf 104}, 121102 (2010)
  [arXiv:1001.3856 [hep-ph]].
  %%CITATION = PRLTA,104,121102;%%



%\cite{Beneke:2010dz}
\bibitem{Beneke:2010dz}
  M.~Beneke, B.~Garbrecht, C.~Fidler, M.~Herranen and P.~Schwaller,
  ``Flavoured Leptogenesis in the CTP Formalism,''
  Nucl.\ Phys.\  B {\bf 843} (2011) 177
  [arXiv:1007.4783 [hep-ph]].
  %%CITATION = NUPHA,B843,177;%%


%\cite{Garny:2010nz}
\bibitem{Garny:2010nz}
  M.~Garny, A.~Hohenegger and A.~Kartavtsev,
  ``Quantum corrections to leptogenesis from the gradient expansion,''
  arXiv:1005.5385 [hep-ph].
  %%CITATION = ARXIV:1005.5385;%%

%\cite{Kolb:1979qa}
\bibitem{Kolb:1979qa}
  E.~W.~Kolb and S.~Wolfram,
  ``Baryon Number Generation In The Early Universe,''
  Nucl.\ Phys.\  B {\bf 172} (1980) 224
  [Erratum-ibid.\  B {\bf 195} (1982) 542].
  %%CITATION = NUPHA,B172,224;%%

%\cite{Buchmuller:2004nz}
\bibitem{Buchmuller:2004nz}
  W.~Buchm\"uller, P.~Di Bari and M.~Pl\"umacher,
  ``Leptogenesis for pedestrians,''
  Annals Phys.\  {\bf 315} (2005) 305
  [arXiv:hep-ph/0401240].
  %%CITATION = APNYA,315,305;%%

%\cite{DiBari:2005st}
\bibitem{DiBari:2005st}
  P.~Di Bari,
  ``Seesaw geometry and leptogenesis,''
  Nucl.\ Phys.\  B {\bf 727} (2005) 318
  [arXiv:hep-ph/0502082].
  %%CITATION = NUPHA,B727,318;%%

%\cite{Engelhard:2006yg}
\bibitem{Engelhard:2006yg}
  G.~Engelhard, Y.~Grossman, E.~Nardi and Y.~Nir,
  ``Importance of the Heavier Singlet Neutrino in Leptogenesis,''
  Phys.\ Rev.\ Lett.\  {\bf 99} (2007) 081802
  [arXiv:hep-ph/0612187].
  %%CITATION = PRLTA,99,081802;%%

%\cite{Bertuzzo:2010et}
\bibitem{Bertuzzo:2010et}
  E.~Bertuzzo, P.~Di Bari and L.~Marzola,
  ``The problem of the initial conditions in flavoured leptogenesis and the
  tauon $N_2$-dominated scenario,''
  arXiv:1007.1641 [hep-ph].
  %%CITATION = ARXIV:1007.1641;%%


%\cite{Endoh:2003mz}
\bibitem{Endoh:2003mz}
  T.~Endoh, T.~Morozumi and Z.~h.~Xiong,
  ``Primordial lepton family asymmetries in seesaw model,''
  Prog.\ Theor.\ Phys.\  {\bf 111} (2004) 123
  [arXiv:hep-ph/0308276].
  %%CITATION = PTPKA,111,123;%%

%\cite{Abada:2006fw}
\bibitem{Abada:2006fw}
  A.~Abada, S.~Davidson, F.~X.~Josse-Michaux, M.~Losada and A.~Riotto,
  ``Flavour Issues in Leptogenesis,''
  JCAP {\bf 0604}, 004 (2006)
  [arXiv:hep-ph/0601083].
  %%CITATION = JCAPA,0604,004;%%

%\cite{Nardi:2006fx}
\bibitem{Nardi:2006fx}
  E.~Nardi, Y.~Nir, E.~Roulet and J.~Racker,
  ``The importance of flavor in leptogenesis,''
  JHEP {\bf 0601}, 164 (2006)
  [arXiv:hep-ph/0601084].
  %%CITATION = JHEPA,0601,164;%%

%\cite{Fong:2010bh}
\bibitem{Fong:2010bh}
  C.~S.~Fong, M.~C.~Gonzalez-Garcia and J.~Racker,
  ``CP Violation from Scatterings with Gauge Bosons in Leptogenesis,''
  arXiv:1010.2209 [hep-ph].
  %%CITATION = ARXIV:1010.2209;%%

%\cite{Kiessig:2010pr}
\bibitem{Kiessig:2010pr}
  C.~P.~Kiessig, M.~Pl\"umacher and M.~H.~Thoma,
  ``Decay of a Yukawa fermion at finite temperature and applications to
  leptogenesis,''
  Phys.\ Rev.\  D {\bf 82} (2010) 036007
  [arXiv:1003.3016 [hep-ph]].
  %%CITATION = PHRVA,D82,036007;%%


\end{thebibliography}
\end{document}